\pdfoutput=1
\documentclass[pre,preprint,twoside,floatfix,superscriptaddress]{revtex4}
\usepackage{amsmath}
\usepackage{extraipa}
\usepackage{empheq}
\usepackage{amsfonts}
\usepackage{graphicx}
\usepackage{overpic}
\usepackage{natbib}

\bibpunct{}{}{,}{s}{,}{,}

\begin{document}
\title{The short--time self diffusion coefficient of a sphere in a suspension of rigid rods}

\author{J.~Guzowski}
\affiliation{Max--Planck--Institut f\"ur Metallforschung, Heisenbergstr.~3,
  D--70569 Stuttgart, Germany}
\affiliation{Institut f\"ur Theoretische und Angewandte Physik, Universit\"at Stuttgart, Pfaffenwaldring 57, D--70569 Stuttgart, Germany}
\affiliation{Institute of Theoretical Physics, Warsaw University, ul.~Ho\.za 69, 00--681 Warsaw, Poland}

\author{B.~Cichocki}
\affiliation{Institute of Theoretical Physics, Warsaw University, ul.~Ho\.za 69, 00--681 Warsaw, Poland}

\author{E.~Wajnryb}
\affiliation{Institute of Fundamental Technological Research, ul.~\'Swi\c{e}tokrzyska 21, 00--049 Warsaw, Poland}

\author{G.~C.~Abade}
\affiliation{Institute of Theoretical Physics, Warsaw University, ul.~Ho\.za 69, 00--681 Warsaw, Poland}

\date{\today}

\begin{abstract}
The short--time self diffusion coefficient of a sphere in a suspension of rigid rods is calculated in first order in the rod volume fraction $\phi$. For low rod concentrations the correction to the Einstein diffusion constant of the sphere due to the presence of rods is a linear function of $\phi$ with the slope $\alpha$ proportional to the equilibrium averaged mobility diminution trace of the sphere interacting with a single freely translating and rotating rod. The two--body hydrodynamic interactions are calculated using the so--called bead model in which the rod of aspect ratio $p$ is replaced by a stiff linear chain of touching spheres. The interactions between spheres are calculated using the multipole method with 
the accuracy controlled by a multipole truncation order and limited only by the computational power. A remarkable accuracy is obtained already for the lowest truncation order, which enables calculations for very long rods, up to $p=1000$. Additionally, the bead model is checked by filling the rod with smaller spheres. This procedure shows that for longer rods the basic model provides reasonable results varying less than $5\%$ from the model with filling. An analytical expression for $\alpha$ as a function of $p$ is derived in the limit of very long rods. We show that in the first order in $1/\log p$ the correction to the Einstein diffusion constant does not depend on the size of the tracer sphere. The higher order corrections depending on the applied model are computed numerically. An approximate expression is provided, valid for a wide range of aspect ratios. 
\end{abstract} 


\maketitle


\section{Introduction.}
Binary colloidal dispersions of hard spheres and rigid rods have been of growing interest recently.\cite{Tracy_1992,Kluijtmans_2000,Kang_2005,Kang_2006,Kang_2007} In the last two years Kang \textit{et al.}\cite{Kang_2005,Kang_2006,Kang_2007} reported on a series of experiments on self diffusion of macromolecular spheres of different sizes in host fd-virus dispersions. In the first work\cite{Kang_2005} the authors studied tracer diffusion in a suspension of freely moving rods. They derived a theoretical expression for the long--time self diffusion coefficient, based on the Smoluchowski equation for a sphere and a rod, neglecting the hydrodynamic interactions. The quantitative agreement with the experiment was obtained only for the large spheres, while for the small spheres the theoretical values of the diffusion constant turned out to be overestimated. This was believed to be due to the growing relevance of the hydrodynamic interactions. In another two papers long time diffusion of small spheres through an entangled rod network was investigated.\cite{Kang_2006,Kang_2007} In the theoretical approach the hydrodynamic interactions were taken into account with the rod network treated as a screening medium and the screening length remaining a free parameter. 

The present work is an attempt to provide a more fundamental description of the hydrodynamic interactions in the studied system. We focus on self diffusion of the sphere for short times in dilute rod suspensions. The diffusion coefficient can be expanded in the rod volume fraction, with the zeroth order term being the Einstein diffusion constant of an isolated sphere and the first order term incorporating hydrodynamic interactions of the sphere with a single, freely moving rod. We provide a detailed description of these interactions applying the so--called bead model, in which the rod is replaced by a linear chain of touching spheres. 

The simplest bead model, in which the beads are treated like point friction sources, has been first developed by Kirkwood and Riseman\cite{Riseman_1948,Riseman_1950} in 1948 and used in calculations of diffusion coefficients of wormlike chains. Yamakawa \textit{et al.}\cite{Yamakawa_1972} took into account the finite size of beads and used the modified Oseen tensor, obtained from replacing each bead by a shell of smaller subunits. Following this methodology, de la Torre \textit{et al.}\cite{Torre_2003} provided a numerical scheme for the so--called shell model, in which the beads are sufficiently small to reproduce the surface of the studied particle. 
However, a full hydrodynamic description, apart from forces and translational motions of the beads, must take into account higher order multipole components as well as lubrication effects arising for configurations near contact. As an example we mention the works of Durlofsky \textit{et al.},\cite{Durlofsky_1987} who included a number of low-level multipoles as well as lubrication corrections and Ladd,\cite{Ladd_1988} who performed computer simulations incorporating multipoles in Cartesian coordinates. The multipole method used here is based on the original solution in spherical coordinates proposed by Schmitz and Felderhof\cite{Felderhof_1976,Schmitz_and_Felderhof_1978,Schmitz_1982,Schmitz2_1982} and further developed by Cichocki and co--workers\cite{Cichocki_1994} and provides very fast numerical convergence. It has already found many applications, for example in the problem of mobilities of spheres conglomerates\cite{Cichocki_1994,Hinsen_1995,Cichocki_2000} and, more recently, transport coefficients in suspensions.\cite{Cichocki_1999,Cichocki_2002,Cichocki_2003}
In the numerical calculations the multipole truncation order $L_{max}$ is introduced and the accuracy of the results can be easily controlled by changing $L_{max}$. 
Also the lubrication corrections have been incorporated in the scheme, providing highly accurate results for the configurations near contact. Cichocki \textit{et al.}\cite{Cichocki_1999} first noted that only the relative motions shall contribute to the correction and proposed a method, which separates out the collective motions. In the foregoing work a new version of the numerical code is used, appropriate for spheres of different radii and thus enabling better imitation of the particle shapes. We study two different kinds of bead models: in the non--filled case the rod is represented by an array of identical, touching spheres on a straight line, while in the filled case, additional smaller spheres fill the gaps in the chain. 

Apart from the numerical calculations, we provide a detailed theoretical description of the sphere--rod hydrodynamic interactions. In the frame of the bead model we analyze the two--body mobility matrix and derive an expression for the sphere mobility trace valid for large interparticle distances. We provide an analytical result for the diffusion coefficient as a function of the rod aspect ratio $p$ in the limit of large $p$. 

In Section II we start with the description of the system and we derive an expression for the diffusion coefficient in first order in the rod volume fraction $\phi$. For low rod concentrations the correction to the Einstein diffusion constant of the sphere due to the presence of rods is a linear function of $\phi$ with the slope $\alpha$ proportional to the equilibrium averaged mobility diminution trace of the sphere interacting with a single freely translating and rotating rod. In Section III we show how the coefficient $\alpha$ can be calculated numerically for the bead model of the rod using the multipole method. In Section IV we present analytical results, valid for long rods and based on the theoretical analysis of the two--particle mobility matrix presented in Appendix A. In Appendix B we derive an asymptotic form of the sphere mobility for large interparticle distances and long rods, which is used to support the numerical calculations presented and discussed in Section V. In Section VI we summarize the results.

\section{System dynamics.}
We consider a system consisting of a hard sphere of radius $a$ and $N$ identical rigid rods of length $L$ and diameter $D$ performing Brownian motion in a fluid of viscosity $\eta$ and volume $V$. The configuration space $X$ of the system is described by the position of the sphere $\mathbf{r}_0$, the centers of the rods $\mathbf{R}_1,\ldots,\mathbf{R}_N$ and the orientations of the rods $\mathbf{u}_1,\ldots,\mathbf{u}_N$. The Brownian motion of the particles can be described by the Smoluchowski equation for the probability distribution $P(X,t)$ for the configuration $X=(\mathbf{r}_0,\mathbf{R}_1,\ldots,\mathbf{R}_N,\mathbf{u}_1,\ldots,\mathbf{u}_N)$ at time $t$:

\begin{equation}
\frac{\partial}{\partial t}P(X,t)= \mathcal{D}P,
\label{smol}
\end{equation}
where $\mathcal{D}$ is the Smoluchowski operator given by

\begin{equation}
\mathcal{D}=\mathbf{\nabla}_X\cdot
\Big[ k_B T \boldsymbol{\mu}(X)\cdot\mathbf{\nabla}_X+
\boldsymbol{\mu}(X)\cdot\mathbf{\nabla}_X \Phi \Big],
\label{Smoluchowski_operator}
\end{equation}
where $k_B$ is the Boltzmann constant, $T$ is the temperature, $\Phi$ is the interaction potential and a shorthand notation has been used, incorporating summation over the particle indeces. The differential operator $\mathbf{\nabla}_X$ is defined by

\begin{equation}
\mathbf{\nabla}_X=(\mathbf{\nabla}_{\mathbf{r}_0},\mathbf{\nabla}_{\mathbf{R}_1},\ldots,\mathbf{\nabla}_{\mathbf{R}_N},\mathbf{\nabla}_{\mathbf{u}_1},\ldots,\mathbf{\nabla}_{\mathbf{u}_N}),
\end{equation}
where $\mathbf{\nabla}_{\mathbf{r}_0},\mathbf{\nabla}_{\mathbf{R}_i}$ denote gradients with respect to the position of the sphere and rod $i$, respectively, and $\mathbf{\nabla}_{\mathbf{u}_i}$ means the gradient in the spherical coordinates on the unit sphere referring to rod $i$. For an explicit form of this operator we refer to the work of Jones.\cite{Jones_2003} The mobility matrix $\boldsymbol{\mu}(X)$ relates the forces $\mathcal{F}$ and torques $\mathcal{T}$ exerted by the particles on the fluid to their translational and rotational velocities $\mathcal{U}$ and $\varOmega$:

\begin{equation}
\begin{pmatrix}
\mathcal{U}\\
\varOmega
\end{pmatrix}
=
\begin{pmatrix}
\boldsymbol{\mu}^{tt} & \boldsymbol{\mu}^{tr}\\
\boldsymbol{\mu}^{rt} & \boldsymbol{\mu}^{rr}
\end{pmatrix}
\begin{pmatrix}
\mathcal{F}\\
\mathcal{T}
\end{pmatrix},
\label{mobility_matrix}
\end{equation}
where $\mathcal{F}=(\mathbf{F}_0,\mathbf{F}_1,\ldots,\mathbf{F}_N)$ and $\mathcal{U}=(\mathbf{U}_0,\mathbf{U}_1,\ldots,\mathbf{U}_N)$ are both $3(N+1)$ dimensional vectors while $\mathcal{T}=(\mathbf{T}_1,\ldots,\mathbf{T}_N)$ and $\varOmega=(\mathbf{\Omega}_1,\ldots,\mathbf{\Omega}_N)$ are $2N$ dimensional vectors. The part of the mobility matrix referring to the sphere $\boldsymbol{\mu}^{tt}_{00}$ is given by the equation

\begin{equation}
\mathbf{U}_0=\boldsymbol{\mu}_{00}^{tt}(X)\cdot\mathbf{F}_0,
\end{equation} 
where the dependence on the configuration of the whole system $X$ has been indicated.
We assume that the system is in equilibrium and concentrate on the Brownian motion of the sphere. We define the short--time self diffusion coefficient as the time derivative of the mean square displacement, in the limit of short times:

\begin{equation}
D_s=\frac{1}{6}\frac{d}{dt}\langle[\Delta\mathbf{r}_0(t)]^2\rangle\vert_{t=0},
\label{Ds_definition}
\end{equation}
where $\langle\ldots\rangle$ denotes the equilibrium average. The expression (\ref{Ds_definition}) can be evaluated using the Smoluchowski equation (\ref{smol}) and some properties of the Smoluchowski operator (\ref{Smoluchowski_operator}). The derivation can be found in the work of Pusey.\cite{Pusey_1991} The result is:

\begin{equation}
D_s=\frac{1}{3}k_BT\langle\text{Tr}\boldsymbol{\mu}_{00}^{tt}\rangle,
\label{tr}
\end{equation}
Next, denoting for simplicity $\boldsymbol{\mu}_{00}=:\boldsymbol{\mu}_{00}^{tt}$, we apply the cluster expansion

\begin{equation}
\boldsymbol{\mu}_{00}(X)=\boldsymbol{\mu}_{00}^{(1)}(\mathbf{r}_0)
+\sum_{i=1}^N\boldsymbol{\mu}_{00}^{(2)}(\mathbf{r}_0,x_i)
+\sum_{i=1}^N\sum_{j<i}\boldsymbol{\mu}_{00}^{(3)}(\mathbf{r}_0,x_i,x_j)+\ldots,
\label{clust}
\end{equation}
where $x_i=(\mathbf{R}_i,\mathbf{u}_i)$ for $i=1,\ldots,N$ and $\boldsymbol{\mu}_{00}^{(1)}(\mathbf{r}_0)$ is the mobility of a single sphere. Assuming  translational invariance and isotropy of the fluid, we have $\boldsymbol{\mu}_{00}^{(1)}(\mathbf{r}_0)=\mu_0\mathbf{1}$, where $\mu_0=1/\zeta_0$ is the mobility of an isolated sphere and $\zeta_0=6 \pi \eta a$ is the Stokes friction coefficient. Assuming low concentration of the rods and using (\ref{tr}) and (\ref{clust}) we arrive at the expansion of $D_s$ in the rod volume fraction $\phi$:

\begin{equation}
D_s=D_0(1-\alpha \phi + o(\phi)),
\label{ds}
\end{equation}
where $D_0=k_BT\mu_0$ is the Einstein diffusion constant, $\phi=vN/V$ with $v$ being the volume of the rod and the coefficient $\alpha$ is proportional to the averaged trace of the two--particle mobility $\boldsymbol{\mu}_{00}^{(2)}$,

\begin{equation}
\alpha=-\frac{1}{3\mu_0 v}\int d\mathbf{R} \, g(\mathbf{R}) \text{Tr}[\boldsymbol{\mu}_{00}^{(2)}(\mathbf{R})],
\label{Alpha}
\end{equation}
where the rod is assumed to be centered at the origin, $\mathbf{R}$ is the sphere position and $g(\mathbf{R})$ is the two--particle low concentration distribution function given by\cite{Hansen_McDonald}

\begin{equation}
g(\mathbf{R})=\left\{
\begin{array}{ll}
0, & \text{sphere and rod overlap}\\
1, & \text{sphere and rod non overlap}
\end{array} \right.
\label{g}
\end{equation}
In the next section we show how the sphere mobility $\boldsymbol{\mu}_{00}$ and, hence, the coefficient $\alpha$ can be calculated numerically using the multipole method. 

\section{The mobility matrix of sphere in presence of rod.}
\begin{figure}
\begin{center}
\includegraphics[width=\textwidth]{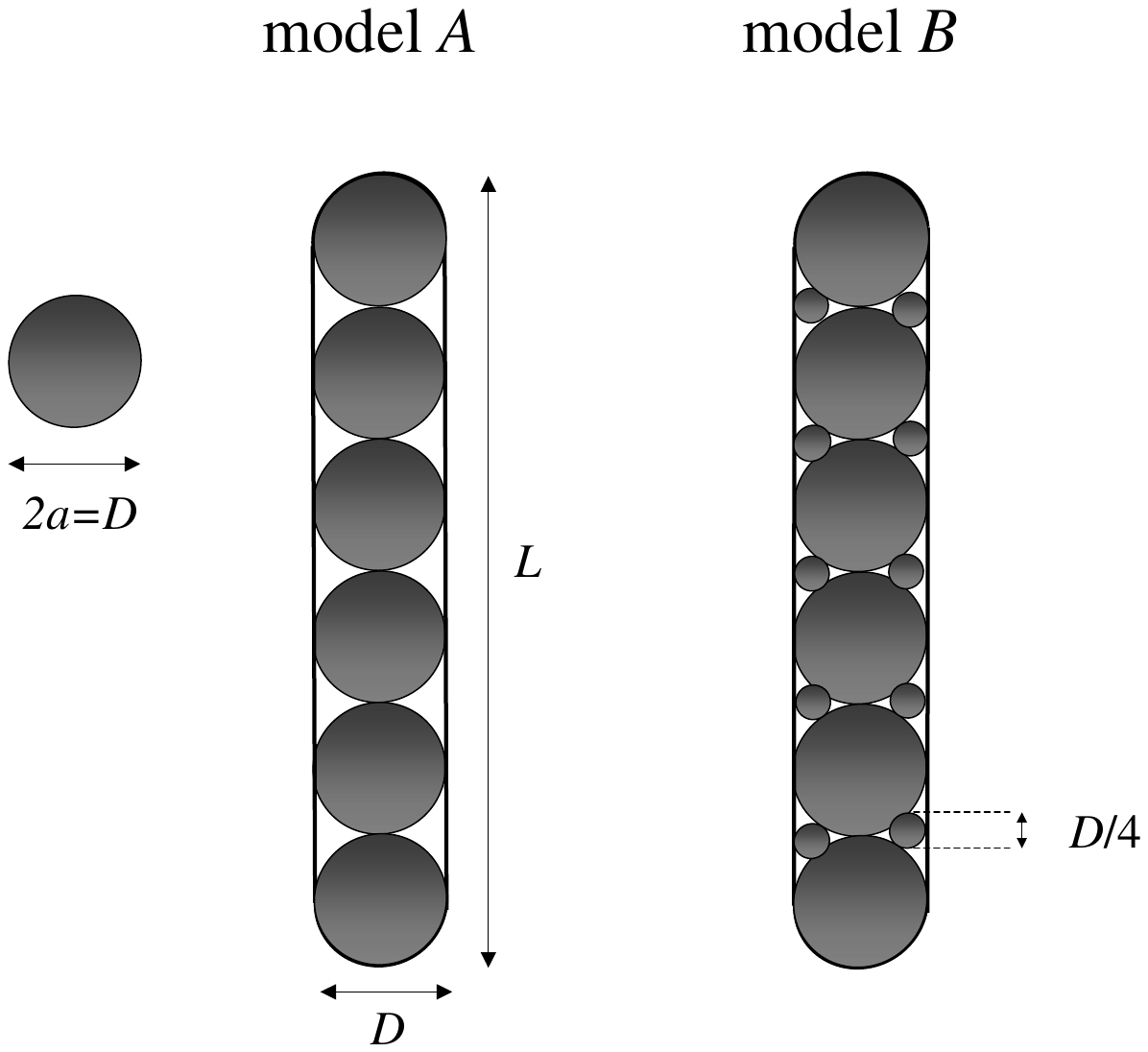}
\end{center}
\caption{Side view of the two different hydrodynamic models of the rod and the tracer sphere (on the left). The small spheres in model B form circular rings (only 2 of 9 spheres in each ring are shown).}
\label{Models}
\end{figure}

For the calculation of the hydrodynamic interactions we replace the rod of aspect ratio $p=L/D$ by a stiff chain of $p$ identical touching spheres (FIG. \ref{Models}, model $A$). 
We propose also an advanced model (FIG. \ref{Models}, model $B$) in which the gaps between the spheres are filled with rings of additional smaller spheres. In the following we briefly describe the multipole method for the hydrodynamic interaction between spheres.

Imagine that we put $N^s$ spheres at positions $(\mathbf{R}^s_1,\ldots,\mathbf{R}^s_{N^s})$, moving with prescribed translational velocities $\mathcal{U}^s=(\mathbf{U}^s_1,\ldots,\mathbf{U}^s_{N^s})$ and rotational velocities $\varOmega^s=(\mathbf{\Omega}^s_1,\ldots,\mathbf{\Omega}^s_{N^s})$, in a viscous unbounded fluid characterised by some incident velocity field $\mathbf{v}_0(\mathbf{r})$. We assume that the resulting fluid flow $\mathbf{v}(\mathbf{r})$ is governed by the Stokes equations,\cite{Kim_Karilla,Happel}

\begin{equation}
-\mathbf{\nabla}\text{p} + \eta \mathbf{\nabla}^2 \mathbf{v}=0,\quad
\mathbf{\nabla} \cdot \mathbf{v}=0, \label{Stokes}
\end{equation} 
where $\text{p}=\text{p}(\mathbf{r})$ is the pressure field. The resulting forces $\mathcal{F}^s=(\mathbf{F}^s_1,\ldots,\mathbf{F}^s_{N^s})$ and torques $\mathcal{T}^s=(\mathbf{T}^s_1,\ldots,\mathbf{T}^s_{N^s})$, exerted by the particles on the fluid, can be related to the particle velocities by the resistance matrix\cite{Schmitz2_1982} $\boldsymbol{\zeta}$

\begin{equation}
\begin{pmatrix}
\mathcal{F}^s\\
\mathcal{T}^s
\end{pmatrix}
=
\begin{pmatrix}
\boldsymbol{\zeta}^{tt} & \boldsymbol{\zeta}^{tr}\\
\boldsymbol{\zeta}^{rt} & \boldsymbol{\zeta}^{rr}
\end{pmatrix}
\begin{pmatrix}
\mathcal{U}^s-v_0\\
\varOmega^s-\omega_0
\end{pmatrix},
\label{Grand_resistance_matrix}
\end{equation}
where:

\begin{gather}
v_0=(\mathbf{v}_0(\mathbf{R}^s_1),\ldots,\mathbf{v
}_0(\mathbf{R}^s_{N^s})), \qquad
\omega_0=(\boldsymbol{\omega}_0(\mathbf{R}^s_1),\ldots,\boldsymbol{\omega}_0(\mathbf{R}^s_{N^s})),
\end{gather}

and

\begin{equation}
\boldsymbol{\omega}_0(\mathbf{r})=\dfrac{1}{2}\nabla\times\mathbf{v}_0(\mathbf{r}).
\end{equation}
The forces $\mathcal{F}^s$ and torques $\mathcal{T}^s$ are determined by the boundary conditions on the particle surfaces. The problem can be solved by introducing force densities\cite{Mazur_1974,Felderhof_1976} $\mathbf{f}_i(\mathbf{r})$ induced on surfaces of each particle $i=1,\ldots,N^s$. For the stick boundary conditions the fluid velocity on a surface $S_i$ of particle $i$ equals

\begin{equation}
\mathbf{w}_i(\mathbf{r})=\mathbf{U}^s_i + \mathbf{\Omega}^s_i\times(\mathbf{r}-\mathbf{R}^s_i)
\end{equation}
and we obtain the following equation on the force densities:

\begin{equation}
\mathbf{w}_i(\mathbf{r}) = \mathbf{v}_0(\mathbf{r})
+ \sum_{j=1}^N
\int d \mathbf{r'} \, \mathbf{T}(\mathbf{r}-\mathbf{r'}) \cdot\mathbf{f}_j(\mathbf{r'}), \qquad \mathbf{r}\in S_i, \qquad i=1,\ldots,N^s
\label{Force_density}
\end{equation}
where $\mathbf{T}(\mathbf{r})$ is the fundamental solution of the Stokes equations (\ref{Stokes}) called the Oseen tensor\cite{Kim_Karilla} and equal to

\begin{equation}
\mathbf{T}(\mathbf{r})=\frac{1}{8\pi\eta r}(\mathbf{1}+\hat{\mathbf{r}}\hat{\mathbf{r}}),
\label{Oseen_tensor}
\end{equation}
with $r=|\mathbf{r}|$ and $\hat{\mathbf{r}}=\mathbf{r}/r$. Once the force densities are known, the total forces and torques can be calculated:

\begin{align}
\mathbf{F}^s_i &= \int d \mathbf{r} \, \mathbf{f}_i(\mathbf{r}),\\
\mathbf{T}^s_i &= \int d \mathbf{r} \, (\mathbf{r}-\mathbf{R}^s_i)\times\mathbf{f}_i(\mathbf{r}),
\end{align}
where $i=1,\ldots,N^s$.
Equation (\ref{Force_density}) can be solved using the multipole expansion in spherical coordinates. By projection onto the complete set of multipole functions\cite{Schmitz_1982} $v_{lm\sigma}(i)$, with $l=1,2,\ldots$, $m=-l,\ldots,l$, $\sigma=0,1,2$ and $i=1,\ldots,N^s$, being the solutions of the Stokes equations (\ref{Stokes}), we obtain an infinite set of linear algebraic equations on the force multipoles $f_{lm\sigma}(i)$. In matrix representation in the multipole space, Eq. (\ref{Force_density}) becomes

\begin{equation}
\mathbf{w}(i)-\mathbf{v}_0(i)=\sum_{j\ne i}\mathbf{G}(ij)\mathbf{f}(j) + \mathbf{Z}_0^{-1}(i)\mathbf{f}(i),\qquad i=1,\ldots,N^s
\label{Integral}
\end{equation}
where the contribution from the force density located on the particle $i$ and contributions from the other particles have been separated. The multipole matrix elements of the single particle friction operator $\mathbf{Z}_0(i)$ and the propagator $\mathbf{G}(ij)$ can be 
found in Refs. 18 and 20. In an abbreviated form, including the summation over the particle indices, the formal solution reads

\begin{equation}
\mathbf{f}=(\mathbf{G}+\mathbf{Z}_0^{-1})^{-1}(\mathbf{w}-\mathbf{v}_0),
\label{Resistance_multipole}
\end{equation}
In the last step the Cartesian components of the forces $\mathbf{F}^s_i$ and torques $\mathbf{T}^s_i$ are expressed by the force multipoles $f_{1m0}(i)$ and $f_{1m1}(i)$, respectively, whereas the Cartesian components of the relative translational velocities $\mathbf{U}^s_i-\mathbf{v}_0(\mathbf{R}_i)$ and the relative rotational velocities $\mathbf{\Omega}^s_i-\boldsymbol{\omega}_0(\mathbf{R}_i)$ are expressed by the velocity multipoles $v_{1m0}(i)$ and $v_{1m1}(i)$, respectively. In other words, the resistance matrix (\ref{Grand_resistance_matrix}) is obtained by a projection of Eq. (\ref{Resistance_multipole}) onto the $FT$--subspace. 
 
In the numerical applications the multipole series is truncated and only the multipoles with  $l\le L_{max}$ are taken into account. The resulting resistance matrix $\boldsymbol{\zeta}_{L_{max}}$ converges quickly with $L_{max}$,\cite{Hinsen_1995} provided there are no relative motions between the spheres. Otherwise, for the configurations near contact, the lubrication effects must be taken into account. This has been done by Cichocki \textit{et. al},\cite{Cichocki_1994} who adapted the lubrication correction to the multipole scheme. It has been shown that separating out the collective motions and applying the lubrication corrections only to the relative motions leads to faster convergence of the multipole series.\cite{Cichocki_2002}

Having obtained the $N^s$--particle resistance matrix we can group the spheres into two rigid assemblies, one being a single sphere and the second composed of remaining spheres and forming a stiff linear chain. The corresponding two--body resistance matrix for sphere and rod is obtained by a linear transformation according to the rigid--body constraints imposed on the spheres in the rod. Finally, the two--body mobility matrix is calculated by an inversion. Taking the translational part referring to the sphere, we obtain the desired quantity $\boldsymbol{\mu}_{00}$.
In the following section we derive an analytical expression for the coefficient $\alpha$ in the limit of large aspect ratios $p$. We show that the leading contribution can be obtained from the bead model taking into account only the lowest multipole.

\section{Analytical results for long rods.}

\subsection{The Oseen approximation.}
In many cases a realistic calculation of hydrodynamic properties of sphere conglomerates requires incorporating a large number of multipoles.\cite{Hinsen_1995} However, it is known that in case of linear chains, built of $p$ spheres, the leading term in $p$ results from the lowest multipole.\cite{Riseman_1950,Cichocki_1994} The torque and rotations as well as higher multipoles are neglected, so that the spheres are treated as point friction sources. The force acting on a chosen bead can be written as a sum of the single particle Stokes drag and the contributions from other beads:  

\begin{equation}
\mathbf{F}^s_i=\zeta^s_i[\mathbf{U}^s_i - \mathbf{v}_0(\mathbf{R}^s_i)-\sum_{j\ne i}^p \mathbf{T}_{ij}\cdot\mathbf{F}^s_j], \qquad i=1,\ldots,p
\label{Oseen}
\end{equation}
where $\zeta^s_i$ is the Stokes friction coefficient of sphere $i$ and $\mathbf{T}_{ij}=\mathbf{T}(\mathbf{R}^s_i-\mathbf{R}^s_j)$. Equation (\ref{Oseen}) is known in the literature as the Oseen approximation.\cite{Oseen_1927,Riseman_1948} In this model the resistance matrix reduces to its translational part denoted by $\boldsymbol{\zeta}_{ij}$, $i,j=1,\ldots,p$:

\begin{equation}
\mathbf{F}^s_i=\sum_{j=1}^p\boldsymbol{\zeta}_{ij}\cdot\mathbf{U}^s_j, \qquad i=1,\ldots,p
\end{equation}
So far we did not make any assumptions for the configuration of the spheres. In the special case of a linear chain, due to the axial symmetry, assuming equal sizes of spheres so that $\zeta^s_i=\zeta$ for all $i$, the resistance matrix $\boldsymbol{\zeta}_{ij}$ can be expressed by two scalar functions $\phi_{ij}$ and $\psi_{ij}$,

\begin{equation}
\boldsymbol{\zeta}_{ij}=\zeta[\phi_{ij}\mathbf{d}\mathbf{d}+\psi_{ij}(\mathbf{1}
-\mathbf{d}\mathbf{d})],
\label{Zeta_ij}
\end{equation}
where $\mathbf{d}$ is the unit vector parallel to the rod axis. If the beads touch each other, we take $\mathbf{R}^s_i-\mathbf{R}^s_j=(i-j)\mathbf{d}$ and the Oseen approximation (\ref{Oseen}) leads to the following equations for the scalar functions:

\begin{align}
\delta_{ij} &= \phi_{ij}+\dfrac{3}{4}\sum_{k=-n, k\ne i}^{k=n}\dfrac{1}{|i-k|}\phi_{kj},\\
\delta_{ij} &= \psi_{ij}+\frac{3}{8}\sum_{k=-n, k \ne i}^{k=n}\dfrac{1}{|i-k|}\psi_{kj},
\end{align}
where the beads are now labeled from $-n$ to $n$ and $2n+1=p$. 
For large $p$ the above discrete equations can be replaced by an integral equation of the form\cite{Riseman_1948}

\begin{gather}
\delta(x-y) = f(x,y)+\lambda\int_{-1}^{1}dtK(x,t)f(t,y),\\
\text{where} \quad K(x,t) = 
\left\{ \begin{array}{lll}
\frac{1}{x-t} & \text{for} & x-t\ge\delta \\
0             & \text{for} & |x-t|<\delta\\
\frac{1}{t-x} & \text{for} & t-x\ge\delta 
\end{array} \right. ,\\
x=\dfrac{i}{n}, \qquad y=\dfrac{j}{n}, \qquad \delta=\dfrac{1}{n},\qquad\\
\text{and}\quad f(x,y) =
\left\{ \begin{array}{lll}
n\phi_{ij} & \text{for} & \lambda=\frac{3}{4}\\
n\psi_{ij} & \text{for} & \lambda=\frac{3}{8} 
\end{array} \right. .\label{f_phi}
\end{gather}

The solution of this type of equation was found by Riseman and Kirkwood\cite{Riseman_1948,Riseman_1950} by means of the Fourier transform. To a detailed discussion of the method we refer also to the review of Zwanzig \textit{et al.}.\cite{Zwanzig_1968} In the limit $p\rightarrow\infty$ we find

\begin{equation}
f(x,y)=\dfrac{1}{2\lambda\log p}\ \delta(x-y) + O((\log p)^{-2}),
\label{Fxy}
\end{equation}
where $\delta(x-y)$ is the Dirac delta. We can apply the Oseen approximation in its general form (\ref{Oseen}) also for a system consisting of an additional sphere outside the chain. This brings us back to the rod--sphere problem. Having found the friction functions $\phi_{ij}$ and $\psi_{ij}$ and using the results from Appendix A, we can calculate the coefficient $\alpha$, defined in Section II, analytically in the limit $p\rightarrow\infty$.

\subsection{Coefficient $\alpha$ in leading order in $p$.}
\begin{figure}
\begin{center}
\includegraphics[width=\textwidth]{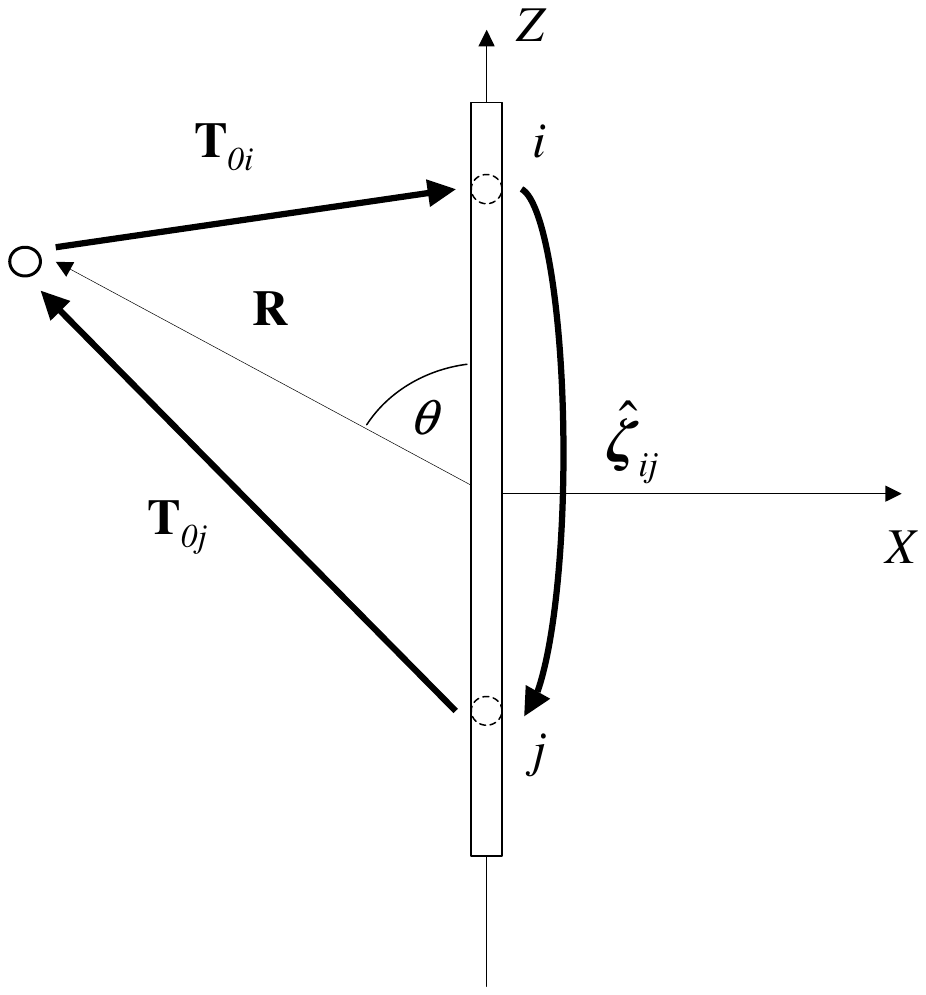}
\end{center}
\caption{Configuration of the sphere interacting with two arbitrary beads $i,j$ in the rod. $\mathbf{T}_{0i}$ denotes the Oseen tensor and $\hat{\boldsymbol{\zeta}}_{ij}$ is the resistance matrix of a freely moving rod. Bold arrows denote consecutive terms in Eq. (\ref{Mu_general}).}
\label{Asymptotic}
\end{figure}

We choose the coordinate system such that the rod axis points in the $Z$ direction and the sphere position $\mathbf{R}$ is described by the distance $R=|\mathbf{R}|$ and the angle $\theta$ between $\mathbf{R}$ and the $Z$--axis (FIG. \ref{Asymptotic}). The mobility matrix of the sphere can be obtained by summing up contributions from all possible scattering sequences between the sphere and the beads in the rod, starting and ending on the sphere. In the limit $p\rightarrow\infty$ the leading contribution comes from the sequence containing two propagators attached to the sphere (FIG. \ref{Asymptotic}). Then:

\begin{equation}
\boldsymbol{\mu}_{00}^{(2)}= -\sum_{ij}\mathbf{T}_{0i}\cdot\hat{\boldsymbol{\zeta}}_{ij}\cdot\mathbf{T}_{0j},
\label{Mu_general}
\end{equation}
where $\hat{\boldsymbol{\zeta}}_{ij}$ is the effective resistance matrix for the beads of the force-- and torque--free rod and 

\begin{equation}
\mathbf{T}_{0i}=\mathbf{T}(\mathbf{R}-\mathbf{d}i).
\end{equation}

Due to the axial symmetry $\hat{\boldsymbol{\zeta}}_{ij}$ can be written in the following form:

\begin{equation}
\hat{\boldsymbol{\zeta}}_{ij}= \zeta[\bar{\phi}_{ij}\mathbf{d}\mathbf{d} + \bar{\psi}_{ij}(\mathbf{1}-\mathbf{d}\mathbf{d})],
\label{Widehat_Zij}
\end{equation}
where $\bar{\phi}_{ij}$ and $\bar{\psi}_{ij}$ are the scalar friction functions modified according to the constraint of rigid body motion of the rod (see Appendix A),

\begin{equation}
\begin{split}
\bar{\phi}_{ij} &= \phi_{ij}-\dfrac{\sum_l\phi_{il}\sum_k\phi_{kj}}{\sum_{kl}\phi_{kl}},\\
\bar{\psi}_{ij} &= \psi_{ij}-\dfrac{\sum_l\psi_{il}\sum_k\psi_{kj}}{\sum_{kl}\psi_{kl}} - \dfrac{\sum_l l\psi_{il}\sum_k k\psi_{kj}}{\sum_{kl} kl\psi_{kl}},
\end{split}
\label{Phipsibar}
\end{equation}
where the functions $\phi_{ij}$ and $\psi_{ij}$ are defined in Eq. (\ref{Zeta_ij}) and we have skipped the summation boundaries. From Eq. (\ref{Phipsibar}) and simple symmetry properties it follows that 

\begin{equation}
\begin{split}
\sum_{i}\bar{\phi}_{ij}&=\sum_{j}\bar{\phi}_{ij}=0,\\ \sum_{i}\bar{\psi}_{ij}&=\sum_{j}\bar{\psi}_{ij}=0,\\  \sum_{i}i\bar{\psi}_{ij}&=\sum_{j}j\bar{\psi}_{ij}=0.
\end{split}\label{Summation}
\end{equation}
We note that the above relations are consistent with the fact that the total force and torque on the rod vanish. 

The expression (\ref{Mu_general}) can be used in the calculation of the coefficient $\alpha$ using Eq. (\ref{Alpha}). 
The Oseen tensors $\mathbf{T}_{0i}$ decay like $O(R^{-1})$, so that the integrals diverge, but the diverging parts cancel under summation. We can avoid dealing with the infinities, when we replace the tensors $\mathbf{T}_{0i}$ by $\mathbf{T}_{0i}-\mathbf{T}(\mathbf{R})$, which behave like $O(R^{-2})$. This can be done as soon as relations (\ref{Summation}) hold. 

We take the rod length $L$ as the length unit and introduce dimensionless quantities

\begin{equation}
\begin{split}
	\mathbf{R}^*              &:=\mathbf{R}/L,\\
	\mathbf{T}^*(\mathbf{R}^*)&:=p\zeta\mathbf{T}(\mathbf{R}),\\
	\hat{\boldsymbol{\zeta}}_{ij}^* &:=\frac{\hat{\boldsymbol{\zeta}}_{ij}}{\zeta}.
\end{split}
\label{Dimensionless}
\end{equation}
The volume $v$ of a cap--ended cylinder of aspect ratio $p$ and diameter $D$ equals

\begin{equation}
	v=\frac{\pi D^2}{4}\left(p-\frac{1}{3}\right).
	\label{Volume}
\end{equation}
In the limit of large rod lengths $L\rightarrow\infty$, keeping the rod diameter $D$ and the tracer sphere radius $a$ constant, the distribution function $g(\mathbf{R}^*)$ equals unity on the whole space and the asymptotic form of $\alpha$, according to Eqs. (\ref{Alpha}), (\ref{Widehat_Zij}), (\ref{Dimensionless}) and (\ref{Volume}), reads:

\begin{equation}
\alpha\xrightarrow[p\rightarrow\infty]{}-\dfrac{4}{3\pi\mu_0\zeta}\sum_{ij}\int d\mathbf{R}^*\text{Tr}\big( [\mathbf{T}^*_{0i}-\mathbf{T}^*(\mathbf{R}^*)]\cdot\hat{\boldsymbol{\zeta}}^*_{ij}\cdot[\mathbf{T}^*_{0j}-\mathbf{T}^*(\mathbf{R}^*)]\big),
\label{Alpha_regularized}
\end{equation} 
Performing the integration leads to 

\begin{equation}
\alpha = -\dfrac{3}{4 p\mu_0\zeta}\sum_{ij}|i-j|\left[\bar{\phi}_{ij} + 3\cdot\bar{\psi}_{ij}\right].
\label{Integrated}
\end{equation}
Using Eqs. (\ref{f_phi}), (\ref{Fxy}) and (\ref{Phipsibar}) and replacing the sums by integrals, we arrive at

\begin{equation}
\alpha=\frac{17}{30\mu_0\zeta}\frac{p}{\log p}[1+O((\log p)^{-1})],
\label{Alpha_Oseen}
\end{equation}
where the higher order corrections $O((\log p)^{-1})$ contain also terms $O(p^{-1})$, which arise from approximating sums by integrals. From Eqs.\ (\ref{Alpha_Oseen}) and (\ref{ds}) it follows, that for very thin rods the first order correction to the Einstein diffusion constant, equal $-D_0\alpha\phi$, is proportional to $L^2/\log p$ and does not depend on the tracer sphere size, as long as the sphere radius $a$ is sufficiently small in comparison with the rod length $L$. More precisely, this result is valid in the limit $\log L\gg \log D$ with $D/a$ kept constant, which means that we must also have $\log L\gg \log a$. However, when this condition is not satisfied, the dependence on $a$ appears in a correction to Eq.\ (\ref{Alpha_Oseen}), which then must be taken into account.




In the next two sections we investigate numerically the case of the tracer sphere equal to the beads in the rod, $2a=D$. Then $\mu_0\zeta=1$ and the prefactor in Eq. (\ref{Alpha_Oseen}) reduces to $17/30$, which we compare with the numerical results for very long rods.

\section{Numerical calculations and discussion of results.}
The integration according to Eq. (\ref{Alpha}) has been performed using the Gaussian quadrature method. Due to the rotational and reflectional symmetry of the mobility matrix, the integration area could be reduced to the first quarter of the $XZ$-plane. 

\begin{figure}
\begin{center}
\begin{overpic}[width=0.8\textwidth]{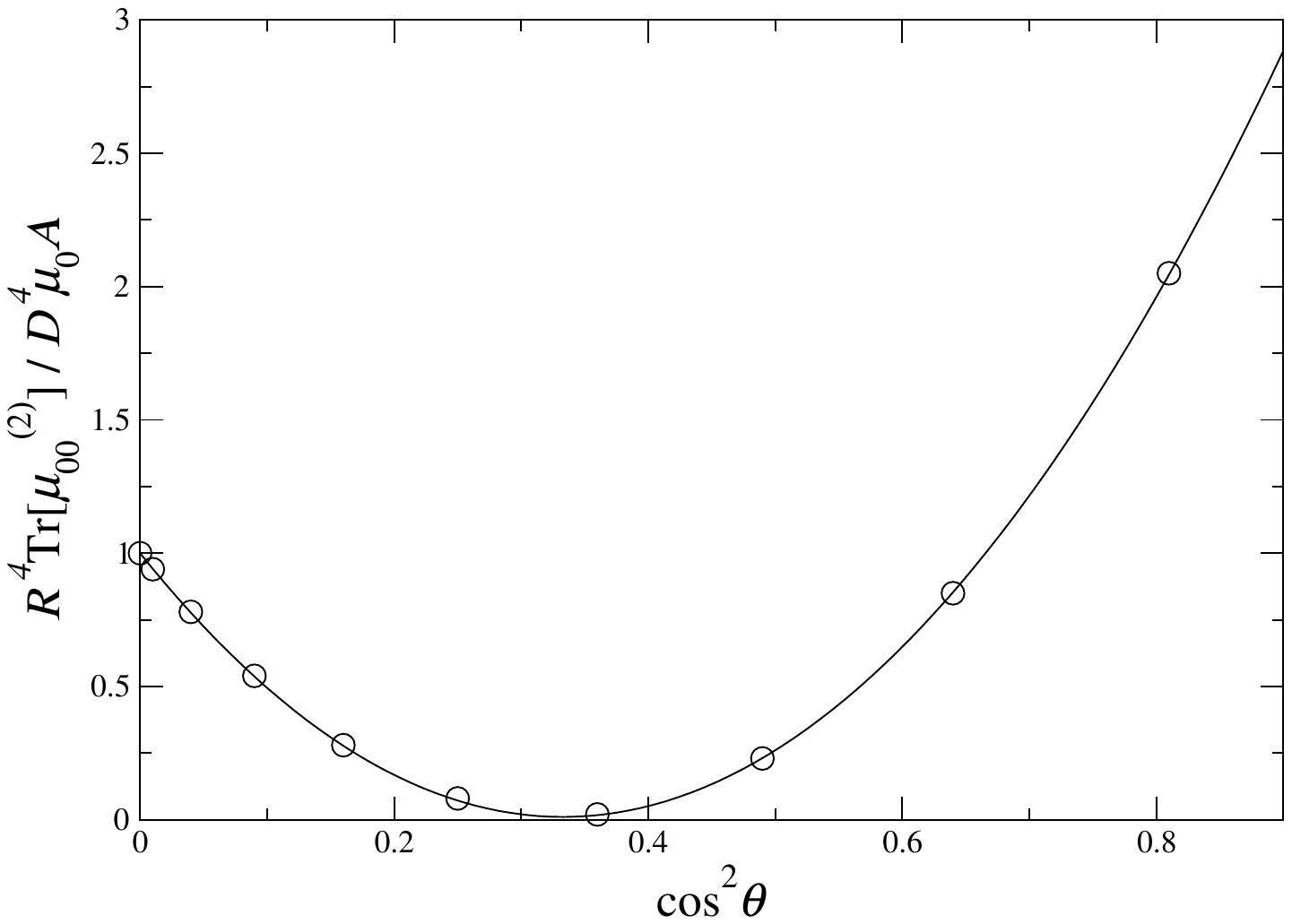}
\put(2,5){$a)$}
\end{overpic}\\
\begin{overpic}[width=0.8\textwidth]{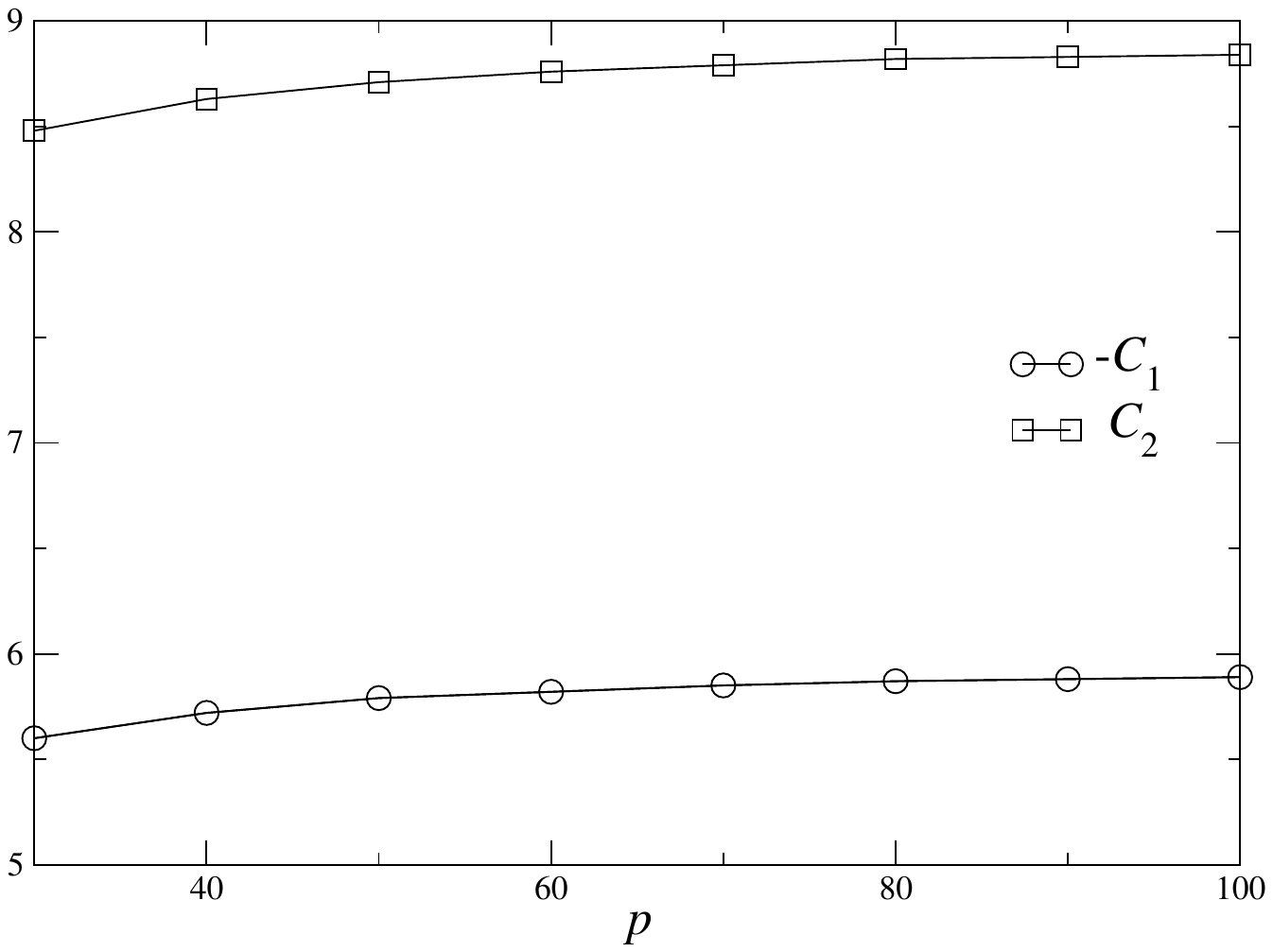}
\put(2,5){$b)$}
\end{overpic}
\end{center}
\caption{a) Angle dependence of the two article mobility trace $\text{Tr}[\boldsymbol{\mu}_{00}^{(2)}]$ normalized to its value for $\theta=\pi/2$. Here $p=100$, $\mathcal{R}=700$ and $L_{max}=1$. Circles are numerical data points and the solid line is a two--parameter fit $1+C_1\cos^2\theta+C_2\cos^4\theta$, see Eq. (\ref{Tr_fit}). b) The plots of the number coefficients $-C_1$ and $C_2$. Theoretical values are 6 and 9, respectively. Solid lines are a guide to the eye.}
\label{C_coefficients}
\end{figure}

\begin{figure}
\begin{center}
\includegraphics[width=\textwidth]{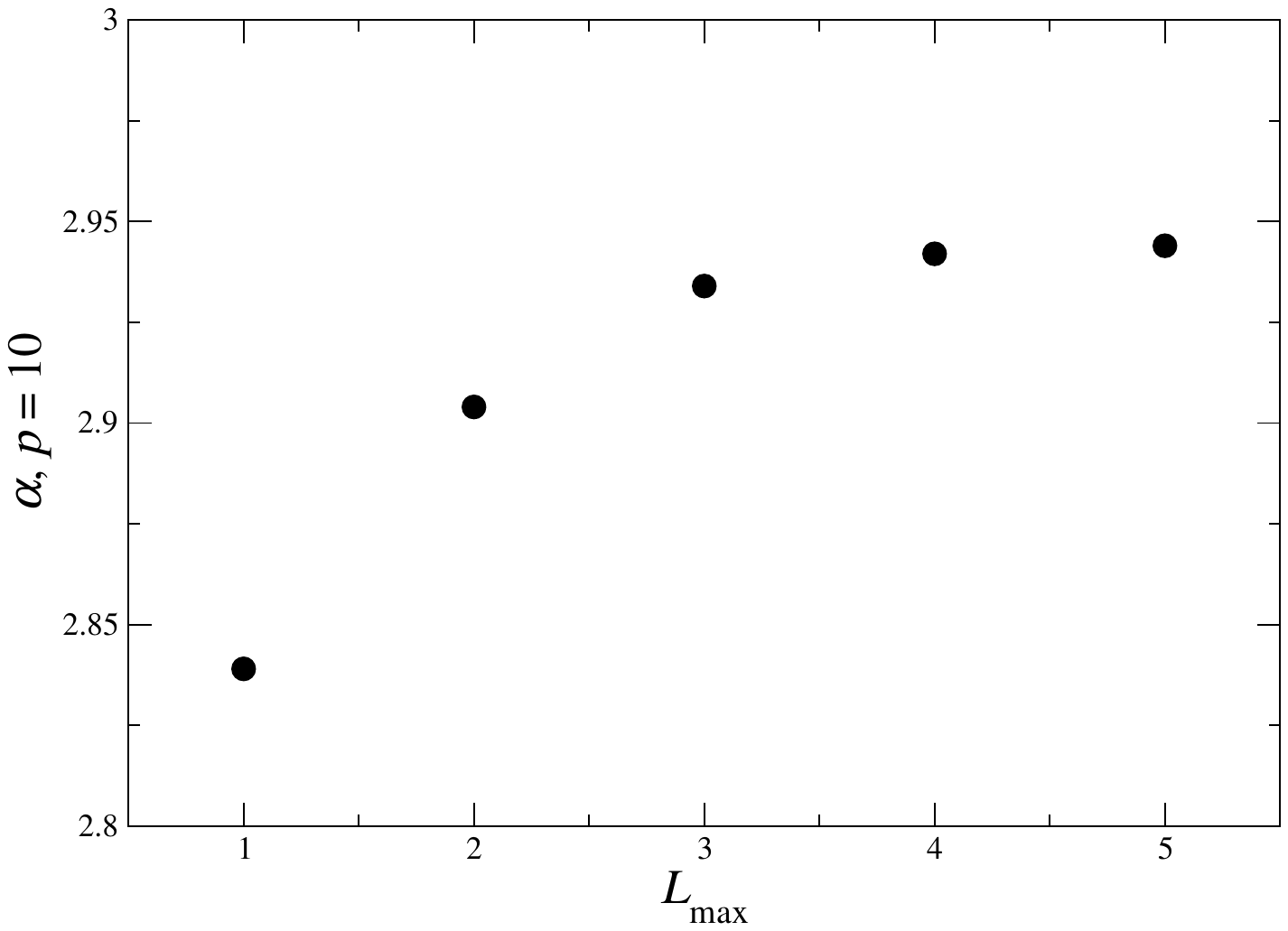}
\end{center}
\caption{The numerical values of the coefficient $\alpha$ as a function of the truncation order $L_{max}$ for $p=10$.}
\label{Fig_Truncation_dependence}
\end{figure}

To optimize the calculations, the numerical integrals have been performed separately on the three sub--areas $A_i$, $i=1,2,3$:

\begin{align*}
A_1&=\{(X,Z):1\le d((X,Z),S)\le X_1\},\\
A_2&=\{(X,Z):X_1\le d((X,Z),S)\le X_2\},\\
A_3&=\{(X,Z):X_2\le d((X,Z),S) \quad \text{and} \quad X^2+Z^2\le (4p)^2\},\\
\end{align*}

where all lengths are normalized to the sphere diameter $D$ and $d((X,Z),S)$ denotes the distance between a point $(X,Z)$ and the rod surface $S$. To optimize the accuracy of the quadratures, $X_i$ have been set such that $X_i/X_{i+1}$ does not exceed 30. 
For the distances $\mathcal{R}=R/D>4p$, according to the considerations in Appendix B, the integral could be performed analytically using the expression:

\begin{equation}
\text{Tr}[\boldsymbol{\mu}_{00}^{(2)}(\mathcal{R},\theta)]=-\frac{\mu_0}{\mathcal{R}^4}
A(1+C_1\cos^2\theta+C_2\cos^4\theta),
\label{Tr_fit}
\end{equation}
where the coefficients $A,C_1,C_2$ were obtained from a fit to the numerical data at $\mathcal{R}=7p$. FIG. \ref{C_coefficients} presents the angle dependence of the two--particle mobility trace normalized to its value for $\theta=\pi/2$ together with the numerical fit, as well as the numerical coefficients as functions of $p$. With growing $p$ the values of $C_1$ and $C_2$ approach -6 and 9, in agreement with the theoretical predictions (\ref{Trace_asymptotic}). 

The accuracy of the results has been found to depend mostly on the truncation order $L_{max}$. FIG. \ref{Fig_Truncation_dependence} shows the results for different $L_{max}$ for $p=10$. It can be seen that for $L_{max}=3$ the error becomes smaller than $0.5\%$ and for $L_{max}=1$ it is still only about $4\%$. 
This fast convergence can be attributed to the following features of the system. Firstly, the spheres in the rod do not exhibit relative motions and the lubrication corrections are not necessary. Secondly, as the spheres are put on a straight line, only the interactions vanishing like $R^{-1}$ are long--ranged (in the sense that their one--dimensional integral diverges). Accordingly, truncation at $L_{max}=1$ gives results converging to those at $L_{max}=3$ with the growing rod length. This is in contrast to the case of three--dimensional conglomerates of spheres studied in Ref. 17, for which also $R^{-2}$ and $R^{-3}$ terms exhibit the long--range character and all multipoles up to $L_{max}=3$ must be incorporated.
 
Due to the limitation of the computer memory, only the values of $p$ up to $10^3$ have been accessible. Then, as long as $\log p\approx 6$ is not a big number, the asymptotic form inEq. (\ref{Alpha_Oseen}) should not be expected to be very accurate. Moreover, the typical experimental values of $p$ are of the order of $10^1$ or $10^2$. Hence, for the quantitative comparison with the numerical data we have introduced an approximate formula incorporating the logarithmic corrections of higher order:

\begin{equation}
\alpha=\frac{17}{30}\frac{p}{\log p - \gamma(p)},
\label{Alpha_Oseen_corrected}
\end{equation}
where $\gamma(p)$ is a function depending on the shape details of the rod. From the theoretical considerations (\ref{Alpha_Oseen}) we can suppose that for large $p$ the function should be approximately given by a series in $(\log p)^{-1}$. From a fit to the numerical data for $20<p<1000$ we have obtained  

\begin{equation}
\gamma(p)\approx 2.12-\dfrac{4.39}{\log p},
\label{Gamma_lmax_1}
\end{equation}
which reproduces the numerical values for $L_{max}=3$ within 2\% accuracy for $p>12$.

\begin{figure}
\begin{center}
\begin{overpic}[width=0.8\textwidth]{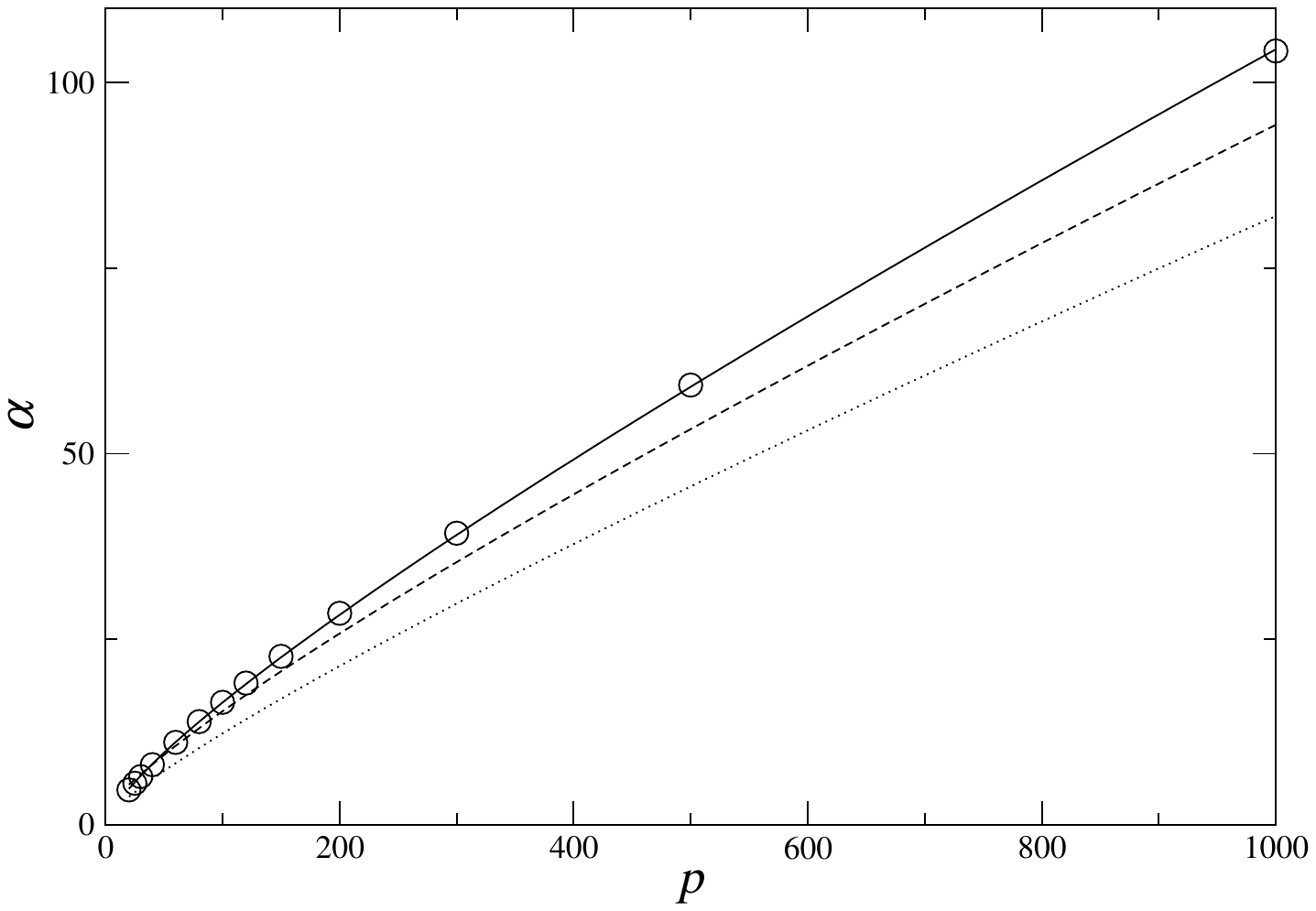}
\put(2,5){$a)$}
\end{overpic}
\begin{overpic}[width=0.8\textwidth]{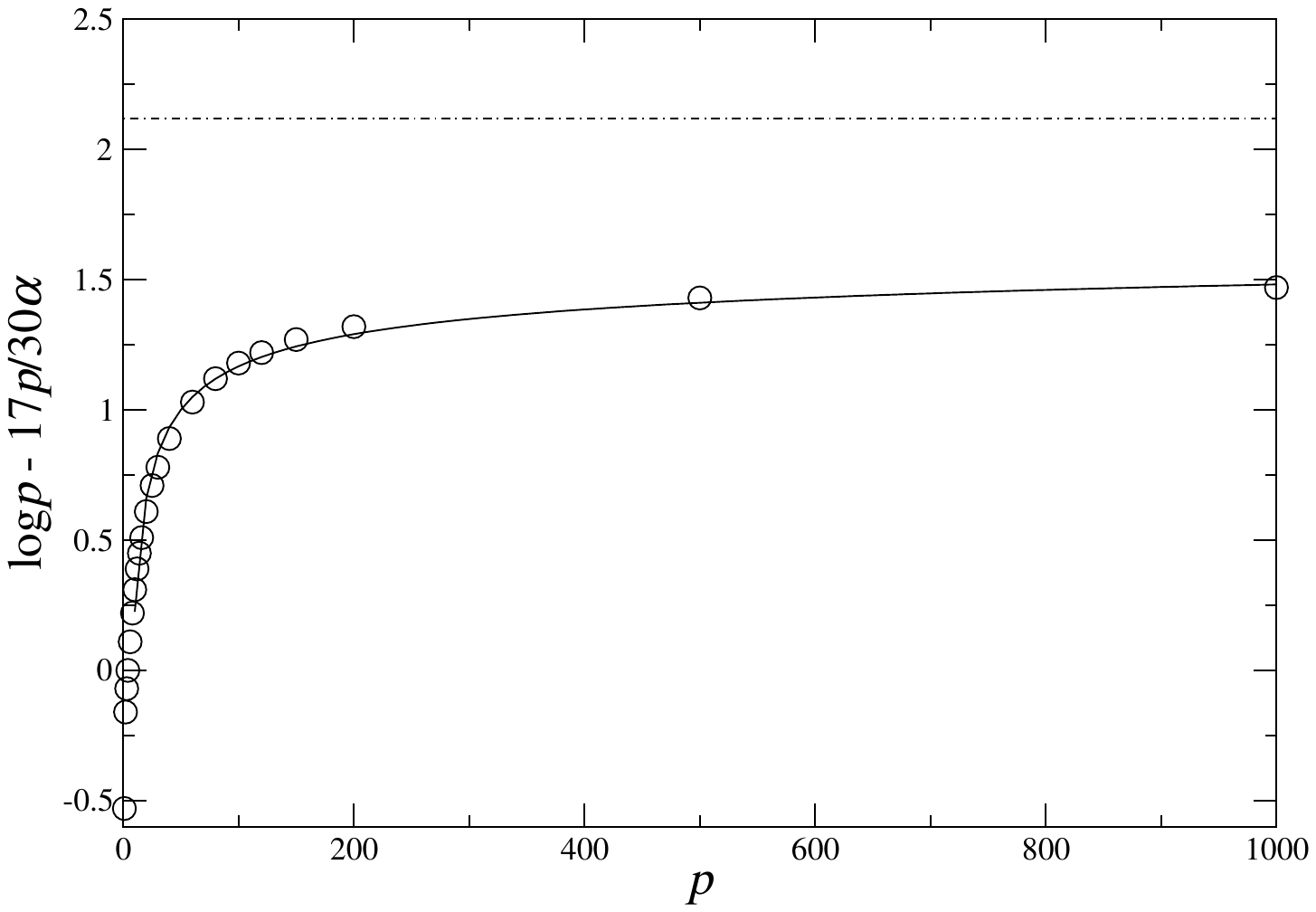}
\put(2,5){$b)$}
\end{overpic}
\end{center}
\caption{
$a)$ The numerical values of the coefficient $\alpha$ as a function of the rod aspect ratio $p$ for the truncation order $L_{max}=1$ (circles) together with the expression $\alpha=17p/30(\log p - \gamma(p))$ for $\gamma(p)=2.12-4.39/\log p$ obtained from a fit to the numerical data for $20<p<1000$ (solid line), $\gamma=0.9$ for the problem of intrinsic viscosity taken from Ref. 9 (dashed line), $\gamma=0$ corresponding to the Oseen approximation (dotted line). $b)$ The numerical values of $\log p -17p/30\alpha$ together with the numerical fit $\gamma(p)$. We note a very slow convergence to the limiting value $2.12$ (marked by the horizontal line).
}
\label{Alpha_long}
\end{figure}

\begin{figure}
\begin{center}
\begin{overpic}[width=1.1\textwidth]{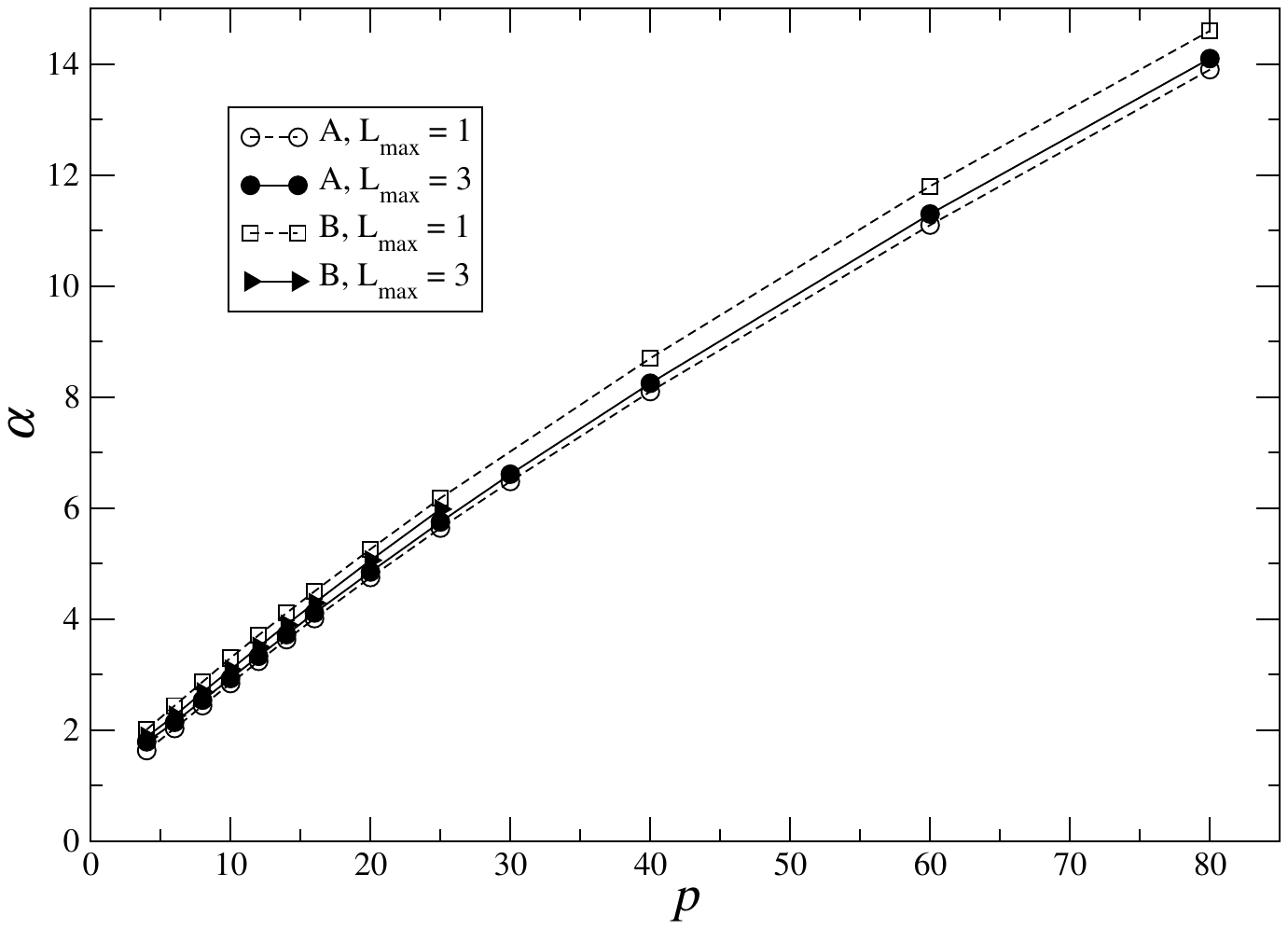}
\put(43,8){\includegraphics[width=0.5\textwidth]{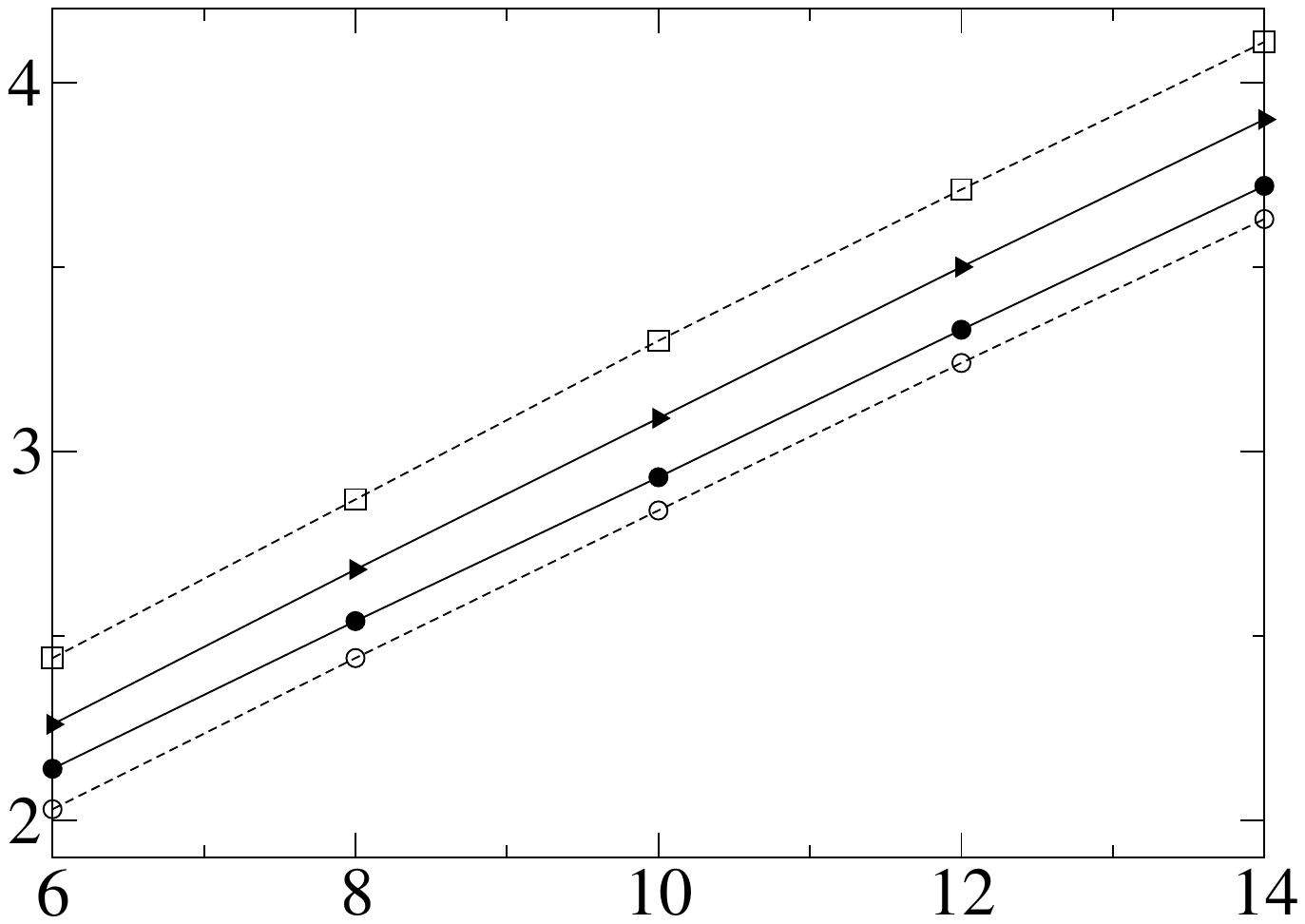}}
\end{overpic}
\end{center}
\caption{The coefficient $\alpha$ as a function of the rod aspect ratio $p$ for the two models of the rod ($A$ and $B$) and the truncation orders $L_{max}=1$ and $L_{max}=3$. In model A the rod is replaced by a stiff chain of equal spheres and in model B additional smaller spheres fill the gaps in the chain (FIG. \ref{Models}). The inset presents the same plots for smaller values of $p$. The lines are guide to the eye. The relative difference between the models diminishes with growing $p$ and is less than 5\% for $p>12$ and $L_{max}=3$. The values converge with growing $L_{max}$, suggesting that incorporating all multipoles would lead to nearly the same results for the two models.}
\label{Alpha_all}
\end{figure}

The numerical values of $\alpha$ as a function of $p$ are presented in FIG. \ref{Alpha_long} and TABLE \ref{Alpha_table}. For $p>20$ the case $L_{max}=1$ differs by less then 2\% from the case $L_{max}=3$. This means that for longer rods it is sufficient to incorporate only multipoles with $l=1$. However, this is still more than in the Oseen approximation, because beside the forces also the higher multipoles $l=1,\sigma=1$ and $l=1,\sigma=2$ are incorporated. As shown in FIG. \ref{Alpha_long} $a)$ the two parameter fit in Eq. (\ref{Gamma_lmax_1}), provides an excellent agreement with the numerical data. We find a deviatiation of about 20\% from the asymptotic form in Eq. (\ref{Alpha_Oseen}), represented by the dotted curve. As mentioned, this is due to the slow convergence of the logarithmic tails. For comparison we present a curve with $\gamma=0.9$, which is the value of a logarithmic correction to the intrinsic viscosity of a suspension of cylinders, obtained numerically by Ortega and Garcia de la Torre\cite{Torre_2003}. 

In FIG. \ref{Alpha_long} $b)$ we compare the expression (\ref{Gamma_lmax_1}) with the numerical values of

\begin{equation}
\log p-\frac{17p}{30\alpha},
\label{Expression}
\end{equation}
Again we find a good agreement between the fit and the data. 

FIG. \ref{Models} presents two different types of bead models applied in the numerical calculations and in FIG. \ref{Alpha_all} and TABLE \ref{Alpha_table} we compare the results. The optimal filling is provided, when the rings consist of 9 spheres of diameter $D/4$. Each of the small spheres touches the two neighboring big spheres and the imaginary rod surface. Accordingly, we deal with 10 times more spheres than in the non--filled case. Because the number of operations grows like $N^3$ with $N$ being the total number of spheres, we were limited in this case to aspect ratios not exceeding $p=25$. The values of $\alpha$ for the filled case diminish with $L_{max}$, opposite to the non--filled case, so that the difference diminishes with growing truncation order. This suggests that our results are very close to the exact values for the cylindrical rod.

\section{Conclusion}

We have calculated the short--time self diffusion coefficient of a hard sphere in a suspension of rigid rods by applying the bead model for the rod and using the highly accurate multipole method for interactions between spheres. 
The numerical values converge very quickly with the truncation order $L_{max}$, giving practically exact results already for $L_{max}=1$ for long rods. This is on the contrary to the case of the drag coefficients of three--dimensional conglomerates of spheres calculated in Ref. 17, where all multipoles up to $L_{max}=3$ must have been taken into account. 
We have shown that the expression for $\alpha$ as a function of the rod aspect ratio $p$ can be derived analytically for large $p$ for the simplest model of spheres treated as point friction sources (see Eq. (\ref{Alpha_Oseen})). In the limit of large rods, the correction to the Einstein diffusion constant has been shown to be independent of the tracer sphere size. 
By applying two different hydrodynamic models for the rod we have checked that only the higher order logarithmic corrections, vanishing like $(\log p)^{-1}$, depend on the rod shape details. For very long rods an approximate, semi--numerical expression for this correction (see Eq. \ref{Gamma_lmax_1}) has been found. The quantitative difference between the models has been found to be less than $5\%$ for $p>12$ with the results converging with growing $p$. This suggests that our results provide bounds for the exact value for the cylindrical rod  and that already the basic bead model (without filling) provides a reasonable approximation. 

So far, we have obtained results for a sphere of a diameter equal to the rod diameter, but the calculations can be performed for tracer spheres of arbitrary sizes. Beside the hard--wall--like repulsions, also the other forms of interparticle potentials can be easily incorporated in the calculations. 

\section*{ACKNOWLEDGMENTS}
This work was supported (EW) by the MANAR Network financed by MNiSzW.The work of G. C. Abade was supported by CAPES Foundation / Ministry of Education of Brazil. One of the authors (J.G.) is indebted to R.Merath, P.Jakubczyk and A.Jajkowska for crictical comments on the manuscript.

\appendix

\section{Mobility matrix for sphere and rod.}
In this appendix we analyze hydrodynamic interactions between a sphere and a rod. In the limit of long rods, using the bead model, we estimate the  behavior of the friction tensors for large $p$ and obtain the mobility matrix $\boldsymbol{\mu}_{00}$ of the sphere in leading order in $p$. Denoting the sphere and the rod by indices $0$ and $1$, respectively,  we define the corresponding two--body resistance matrix in the Cartesian coordinates:\cite{Kim_Karilla}

\begin{equation}
\begin{pmatrix}
\mathbf{F}_0\\
\mathbf{F}_1\\
\mathbf{T}_0\\
\mathbf{T}_1
\end{pmatrix}
=
\begin{pmatrix}
\mathbf{A}_{00} & \mathbf{A}_{01} & \widetilde{\mathbf{B}}_{00} & \widetilde{\mathbf{B}}_{01}\\
\mathbf{A}_{10} & \mathbf{A}_{11} & \widetilde{\mathbf{B}}_{10}  & \widetilde{\mathbf{B}}_{11}\\
\mathbf{B}_{00} & \mathbf{B}_{01} & \mathbf{C}_{00} & \mathbf{C}_{01}\\
\mathbf{B}_{10} & \mathbf{B}_{11} & \mathbf{C}_{10} & \mathbf{C}_{11}\\
\end{pmatrix}
\begin{pmatrix}
\mathbf{U}_0\\
\mathbf{U}_1\\
\mathbf{\Omega}_0\\
\mathbf{\Omega}_1
\label{ABC_matrix}
\end{pmatrix},
\end{equation}
where $\mathbf{A}_{ij}$, $\mathbf{B}_{ij}$, $\widetilde{\mathbf{B}}_{ij}$, $\mathbf{C}_{ij}$, $i,j=1,2$ are the two--particle friction tensors. From the Lorentz reciprocal theorem follow the symmetry properties:

\begin{align}
\mathbf{A}_{ij}&=\mathbf{A}_{ji}^T, & \widetilde{\mathbf{B}}_{ij}&=\mathbf{B}_{ji}^T, & \mathbf{C}_{ij}&=\mathbf{C}_{ji}^T.
\label{Lorentz}
\end{align}
In the mobility problem we are interested in an inverse relation,

\begin{equation}
\begin{pmatrix}
\mathbf{U}_0\\
\mathbf{U}_1\\
\mathbf{\Omega}_0\\
\mathbf{\Omega}_1
\end{pmatrix}
=
\begin{pmatrix}
\mathbf{a}_{00} & \mathbf{a}_{01} & \widetilde{\mathbf{b}}_{00} & \widetilde{\mathbf{b}}_{01}\\
\mathbf{a}_{10} & \mathbf{a}_{11} & \widetilde{\mathbf{b}}_{10}  & \widetilde{\mathbf{b}}_{11}\\
\mathbf{b}_{00} & \mathbf{b}_{01} & \mathbf{c}_{00} & \mathbf{c}_{01}\\
\mathbf{b}_{10} & \mathbf{b}_{11} & \mathbf{c}_{10} & \mathbf{c}_{11}\\
\end{pmatrix}
\begin{pmatrix}
\mathbf{F}_0\\
\mathbf{F}_1\\
\mathbf{T}_0\\
\mathbf{T}_1
\label{ABC_mobility}
\end{pmatrix},
\end{equation}
where the mobility tensors $\mathbf{a}_{ij}$, $\mathbf{b}_{ij}$, $\widetilde{\mathbf{b}}_{ij}$, $\mathbf{c}_{ij}$, $i,j=1,2$ have symmetry properties analogical to those for the friction tensors in Eq. (\ref{Lorentz}). We note that $\mathbf{a}_{00}=\boldsymbol{\mu}_{00}$ is the desired mobility matrix of the sphere in presence of a rod. First we estimate the elements of the resistance matrix using the bead model for large $p$ and keep only the leading terms. Like in Section IV, we introduce a dimensionless distance $R^*=R/L$ and take the limit $p\rightarrow\infty$ with $R^*=const$. 

Each of the tensors in Eq. (\ref{ABC_matrix}) refers to a particular physical situation. As an example we choose $\mathbf{A}_{01}$, which describes the case of a translating rod and a fixed sphere.
The fluid velocity field, produced by the rod moving with translational velocity $\mathbf{U}_1$, results in the force $\mathbf{A}_{01}\mathbf{U}_1$, exerted by the sphere on the fluid. In the bead model this force is given by a sum over the beads $i=1,\ldots,p$ of a product $\zeta_0\mathbf{T}_{0i}\mathbf{F}^s_i$, where $\mathbf{T}_{0i}$ is the Oseen tensor  and $\mathbf{F}^s_i$ is the force on bead $i$. The latter is again related to the velocities of the beads by the resistance matrix $\boldsymbol{\zeta}_{ij}$. Taking all velocities equal to $\mathbf{U}_1$, we finally obtain:

\begin{equation}
\mathbf{A}_{01}= \zeta_0\sum_{ij}\mathbf{T}_{0i}\cdot\boldsymbol{\zeta}_{ij},
\label{Example}
\end{equation}
where further reflections between the tracer sphere and the beads can be neglected when $p\rightarrow\infty$. To estimate the asymptotic dependence of $\mathbf{A}_{01}$ on $p$, we use the axisymmetric form (\ref{Zeta_ij}) of $\boldsymbol{\zeta}_{ij}$ with the friction functions $\phi_{ij}$ and $\psi_{ij}$. In the limit $p\rightarrow\infty$, according to Eq. (\ref{Fxy}), the functions differ only by a prefactor, so it is sufficient to estimate the double sum $\sum_{ij}\phi_{ij}/R^*_{0i}$, where the factor $1/R^*_{0i}$ comes from the Oseen tensor and $R_{0i}^*=R_{0i}/L$. Using the integral representation, we obtain

\begin{equation}
\sum_{ij}\dfrac{\phi_{ij}}{R_{0i}}=\frac{1}{R^*}\int_{-1}^{1}d x\int_{-1}^{1}d y \, t(x;R^*,\theta) \, f(x,y),
\label{Estimate}
\end{equation}
where $t(x;R^*,\theta)=R/R_{0i}$ is a function that parametrically depends on $R^*$ and $\theta$ but not on $p$ and the function $f(x,y)$, given by Eq. (\ref{Fxy}), is $O(1/\log p)$. Hence, we we can write Eq. (\ref{Example}) as

\begin{equation}
\mathbf{A}_{01}= \frac{\zeta_0}{\log p}\mathbf{\mathcal{A}}_{01}+o(1/\log p),
\end{equation}
where $\mathbf{\mathcal{A}}_{01}=\mathbf{\mathcal{A}}_{01}(\mathbf{R}^*)$ is a matrix independent of $p$, which we specify at the end of the appendix. Similarly, we obtain

\begin{equation}
	\begin{split}
		\mathbf{A}_{00}&=\zeta_0\mathbf{1}+
			\frac{\zeta_0^2}{\zeta p\log p}\mathbf{\mathcal{A}}_{00}+o(1/p\log p)\\
		\mathbf{A}_{11}&=\frac{\zeta p}{\log p}\mathbf{\mathcal{A}}_{11}+
			o(p/\log p)\\	
		\widetilde{\mathbf{B}}_{01}&=\frac{\zeta_0 p}{\log p}\mathbf{\mathcal{B}}+
			o(p/\log p)\\						
		\mathbf{C}_{11}&=\frac{\zeta p^3}{\log p}\mathbf{\mathcal{C}}+
			o(p^3/\log p)\\				
	\end{split}
\end{equation}
Remaining tensors of the resistance matrix in Eq. (\ref{ABC_matrix}) are of the following orders:

\begin{equation}
\begin{split}
\mathbf{B}_{00}&=O(1/p^2\log p),\\
\mathbf{B}_{01}&=O(1/p\log p),\\
\mathbf{B}_{11}&=O(p/(\log p)^2),\\
\mathbf{C}_{00}&=O(1),\\
\mathbf{C}_{10}&=O(1/\log p).
\end{split}
\end{equation}
Now let us go back to the mobility problem from Eq. (\ref{ABC_mobility}). We can calculate the mobility tensor $\mathbf{a}_{00}$ by putting $\mathbf{F}_1=\mathbf{T}_1=\mathbf{T}_0=0$ in Eq. (\ref{ABC_matrix}) and solving with respect to $\mathbf{F}_0$. We obtain a direct relation between $\mathbf{F}_0$ and $\mathbf{U}_0$, which we invert and keep the leading terms in $p$. Finally, the two particle mobility of the sphere reads

\begin{equation}
\begin{split}
\boldsymbol{\mu}^{(2)}_{00}&=\frac{1}{\zeta p\log p}[\mathbf{\mathcal{A}}_{00} 
-\mathbf{\mathcal{A}}_{01} \cdot \mathbf{\mathcal{A}}_{11}^{-1} \cdot \mathbf{\mathcal{A}}_{01}^T
-\mathbf{\mathcal{B}} \cdot \mathbf{\mathcal{C}}^{-1} \cdot \mathbf{\mathcal{B}}^T] + o(1/p\log p),
\end{split}
\label{Mu_Final}
\end{equation}
where

\begin{equation}
	\begin{split}
		\mathbf{\mathcal{A}}_{00}&=\lim_{p\rightarrow\infty}\frac{\log p}{p}
			\sum_{ij}[\phi_{ij}\mathbf{T}^*_{0i} \cdot \mathbf{d}\mathbf{d} \cdot \mathbf{T}^*_{0j}+
				\psi_{ij}\mathbf{T}^*_{0i} \cdot (\mathbf{1}
				-\mathbf{d}\mathbf{d}) \cdot \mathbf{T}^*_{0j}]\\ 	
		\mathbf{\mathcal{A}}_{01}&=\lim_{p\rightarrow\infty}\frac{\log p}{p}
			\sum_{ij}[\phi_{ij}\mathbf{T}^*_{0i} \cdot \mathbf{d}\mathbf{d}+
				\psi_{ij}\mathbf{T}^*_{0i} \cdot (\mathbf{1}-\mathbf{d}\mathbf{d})]\\
		\mathbf{\mathcal{A}}_{11}&=\lim_{p\rightarrow\infty}\frac{\log p}{p}
			\sum_{ij}[\phi_{ij}\mathbf{d}\mathbf{d}+
				\psi_{ij}(\mathbf{1}-\mathbf{d}\mathbf{d})]\\
		\mathbf{\mathcal{B}}&=\lim_{p\rightarrow\infty}\frac{\log p}{p^2}
			\sum_{ij}i\psi_{ij}\mathbf{T}^*_{0j} \cdot \mathbf{L}_A^T\\
		\mathbf{\mathcal{C}}&=\lim_{p\rightarrow\infty}\frac{\log p}{p^3}
			\sum_{ij}ij\psi_{ij}(\mathbf{1}-\mathbf{d}\mathbf{d}),
	\end{split}
\label{Tensors_d_ij}
\end{equation}
where $\mathbf{L}_A$ is an antisymmetric matrix defined in Eq. (\ref{TensorL}) and $\mathbf{T}_{0i}^*$ is the Oseen tensor normalized to $\zeta p$ (see Eq. (\ref{Dimensionless})).

From Eqs. (\ref{Mu_Final}) and (\ref{Tensors_d_ij}), after straightforward manipulations, we obtain Eqs. (\ref{Mu_general}), (\ref{Widehat_Zij}) and (\ref{Phipsibar}) in the main text.

\section{Limit of large interparticle distances: rod in linear field}
In this Appendix we derive a far--field form of the sphere mobility matrix $\boldsymbol{\mu}_{00}$. Taking the limit $p\rightarrow\infty$ we calculate the coefficients $A,C_1,C_2$ for the two--particle mobility trace in Eq. (\ref{Tr_fit}).

If we assume, that the rod acts on the fluid by the force $\mathbf{F}_1$, the torque $\mathbf{T}_1$ and the symmetric dipole moment $\mathbf{S}_1$, the resulting  velocity field $\mathbf{v}_1(\mathbf{R})$ at a position $\mathbf{R}$ relative to the rod center, where $|\mathbf{R}|>>L$, can be written in the form of the Taylor series\cite{Kim_Karilla}

\begin{equation}	\mathbf{v}_1(\mathbf{r})=\mathbf{F}_1\cdot\mathbf{T}(\mathbf{r})+\mathbf{T}_1\cdot[\frac{1}{2}\nabla\times\mathbf{T}(\mathbf{r})]+\mathbf{S}_1:[\nabla\mathbf{T}(\mathbf{r})]^S+\ldots,
\label{Multipole_rod}
\end{equation}
where the superscript $S$ denotes the symmetric traceless part and

\begin{equation}
	\begin{split}
		\mathbf{F}_1&=\oint_{\partial v} ds \, \mathbf{f}(\mathbf{r}),\\
  	\mathbf{T}_1&=\oint_{\partial v} ds \, \mathbf{r}\times\mathbf{f}(\mathbf{r}),\\	
   	\mathbf{S}_1&=\oint_{\partial v} ds \, [\mathbf{r}\mathbf{f}(\mathbf{r})]^S,			
  \end{split}
\label{Moments_rod}
\end{equation}
where $\mathbf{f}(\mathbf{r})$ is the surface force density on the rod surface $\partial v$.
As soon as the rod moves freely, the lowest non--vanishing force multipole moment is the symmetric dipole moment $\mathbf{S}_1$. In the following, we evaluate $\mathbf{S}_1$ in the limit $p\rightarrow\infty$ using the bead model. 

Consider a single rod immersed in a linear field characterised by a constant vector $\mathbf{u}_{10}$, vorticity $\boldsymbol{\omega}_{10}$ and rate of strain $\mathbf{e}_{10}$. Assume the rod to move freely, such that the total force and torque vanish. The symmetric force dipole moment of the rod $\mathbf{S}_1$ is then given by the following set of equations:

\begin{equation}
\begin{pmatrix}
\mathbf{0}\\
\mathbf{0}\\
\mathbf{S}_1\\
\end{pmatrix}
= 
\begin{pmatrix}
\mathbf{A} & \mathbf{0} & \mathbf{0}\\
\mathbf{0} & \mathbf{C}       & \widetilde{\mathbf{H}}\\
\mathbf{0} & \mathbf{H}        & \mathbf{M} 
\end{pmatrix}
\begin{pmatrix}
\mathbf{U}_1-\mathbf{u}_{10}\\
\mathbf{\Omega}_1-\boldsymbol{\omega}_{10}\\
-\mathbf{e}_{10}
\end{pmatrix},
\label{Rod_friction}
\end{equation} 
where $\mathbf{A},\mathbf{C},\mathbf{H}$ and $\mathbf{M}$ are the friction tensors of the rod and the remaining tensors are zero due to the axial symmetry. According to the Lorentz reciprocal theorem the following relations hold:

\begin{align}
A_{\alpha\beta}&=A_{\beta\alpha},& C_{\alpha\beta}&=C_{\beta\alpha},& H_{\alpha\beta\mu}&= \widetilde{H}_{\mu\alpha\beta},& M_{\alpha\beta\mu\nu}&=M_{\mu\nu\alpha\beta}. 
\end{align}
Due to the rotational invariance around the rod axis the general forms of the friction tensors can be written with use of the unit vector $\mathbf{d}$ and a few scalar coefficients. Following the notation of Kim and Karilla\cite{Kim_Karilla}, they read

\begin{equation}
\begin{split}
&A_{\alpha\beta}=X^Ad_{\alpha}d_{\beta} + Y^A(\delta_{\alpha\beta}-d_{\alpha}d_{\beta}),\\
&C_{\alpha\beta}=X^Cd_{\alpha}d_{\beta} + Y^C(\delta_{\alpha\beta}-d_{\alpha}d_{\beta}),\\
&H_{\alpha\beta\mu}=Y^H\overbracket{d_{\alpha}\epsilon_{\beta\mu\sigma}} {}^{\substack{(\alpha\beta) \ {} \\ {}}}d_{\sigma},
\\
&M_{\alpha\beta\mu\nu}=X^Md_{\alpha\beta\mu\nu}^{(0)}+Y^Md_{\alpha\beta\mu\nu}^{(1)}+Z^Md_{\alpha\beta\mu\nu}^{(2)},
\end{split}
\label{Tensors}
\end{equation}
where the rod friction coefficients depend on the shape of the body and the operation $\overbracket{\qquad} {}^{\substack{(\alpha\beta) \ {} \ {}}}$ means taking the traceless part, symmetric in the indeces $\alpha$ and $\beta$. Explicit forms of the tensors $\mathbf{d}^{(0)}$, $\mathbf{d}^{(1)}$ and $\mathbf{d}^{(2)}$ read

\begin{equation}
\label{d_tensors}
\begin{split}
d_{\alpha\beta\mu\nu}^{(0)}&=\overbracket{d_{\alpha}d_{\beta}} {}^{\substack{(\alpha\beta) \ {} \\ {}}} \overbracket{d_{\mu}d_{\nu}} {}^{\substack{(\mu\nu) \ {} \\ {}}},\\
d_{\alpha\beta\mu\nu}^{(1)}&=\overbracket{ \overbracket{\delta_{\alpha\mu}d_{\beta}d_{\nu}} {}^{\substack{(\alpha\beta) \ {} \\ {}}} } {}^{\substack{(\mu\nu) \ {} \\ {} \\ {}}} - \overbracket{d_{\alpha}d_{\beta}} {}^{\substack{(\alpha\beta) \ {} \\ {}}} \overbracket{d_{\mu}d_{\nu}} {}^{\substack{(\mu\nu) \ {} \\ {}}}, \\
d_{\alpha\beta\mu\nu}^{(2)}&=\overbracket{ \overbracket{\delta_{\alpha\mu}\delta_{\beta\nu}} {}^{\substack{(\alpha\beta) \ {} \\ {}}} } {}^{\substack{(\mu\nu) \ {} \\ {} \\ {}}} -2\overbracket{ \overbracket{\delta_{\alpha\mu}d_{\beta}d_{\nu}} {}^{\substack{(\alpha\beta) \ {} \\ {}}} } {}^{\substack{(\mu\nu) \ {} \\ {}, \\ {}}}+\dfrac{1}{2}\overbracket{d_{\alpha}d_{\beta}} {}^{\substack{(\alpha\beta) \ {} \\ {}}} \overbracket{d_{\mu}d_{\nu}} {}^{\substack{(\mu\nu) \ {} \\ {}}}.
\end{split}
\end{equation}
Solving (\ref{Rod_friction}) with respect to $\mathbf{S}_1$, we obtain

\begin{equation} 
\mathbf{S}_1=-\big[\mathbf{M}-\mathbf{H}\cdot\mathbf{C}^{-1}\cdot\widetilde{\mathbf{H}}\big]:\mathbf{e}_{10}=
-\widehat{\mathbf{M}}:\mathbf{e}_{10},
\label{Stresslet}
\end{equation} 
%
%
where we have introduced a fourth rank tensor $\widehat{\mathbf{M}}$. In the two--body problem we are interested in the velocity field produced by the sphere exerting some prescribed force $\mathbf{F}_0$ on the fluid. The rate of strain around the rod is then given by $\mathbf{e}_{10}=[\nabla\mathbf{T}(\mathbf{r})|_{\mathbf{r}=\mathbf{R}}]^S\cdot\mathbf{F}_0$. 
Accordingly, using Eqs. (\ref{Multipole_rod}), (\ref{Stresslet}) and the conditions $\mathbf{F}_1=0$ and $\mathbf{T}_1=0$, the mobility matrix of the sphere reads

\begin{equation}
\boldsymbol{\mu}_{00}=\mu_0\mathbf{1}-[\nabla\mathbf{T}(\mathbf{r})|_{\mathbf{r}=\mathbf{R}}]^S:\widehat{\mathbf{M}}:[\nabla\mathbf{T}(\mathbf{r})|_{\mathbf{r}=\mathbf{R}}]^S.
\label{Mobility_asymptotic}
\end{equation}
Using the explicit forms in Eqs. (\ref{Tensors}) and (\ref{d_tensors}) we get

\begin{equation} 
\widehat{\mathbf{M}}=X^M\mathbf{d}^{(0)}+\left[Y^M-\frac{(Y^H)^2}{2Y^C}\right]
\mathbf{d}^{(1)}+Z^M\mathbf{d}^{(2)}.
\label{Widehat_M}
\end{equation}
From Eqs. (\ref{Widehat_M}) and (\ref{Mobility_asymptotic}) the angle dependence of the two--particle mobility trace $\text{Tr}[\boldsymbol{\mu}_{00}^{(2)}]$ can be shown to be a linear combination of unity, $\cos^2\theta$ and $\cos^4\theta$. This is due to the form of tensors $\mathbf{d}^{(i)}$, which consist only 2 or 4 vectors $\mathbf{d}$. The distance dependence is $O(R^{-4})$ because of the two derivatives of the Oseen tensor in Eq. (\ref{Mobility_asymptotic}). From these considerations follows the general form in Eq. (\ref{Tr_fit}), where the values of the coefficients $A,C_1,C_2$ can be found in the limit $p\rightarrow\infty$, once the rod friction coefficients $X^M,Y^M,Y^H,Y^C$ and $Z^M$ are calculated. For this purpose we apply the bead model. The individual forces $\mathbf{F}^s_i$ acting on the beads contribute to the total moments and, in the Oseen approximation, we obtain

\begin{equation}
\label{Total_moments}
\begin{split}
&\mathbf{F}_1=\sum_i\mathbf{F}^s_i,\\
&\mathbf{T}_1=
D\sum_ii\mathbf{L}_A\cdot\mathbf{F}^s_i,\\
&\mathbf{S}_1=
D\sum_ii\mathbf{L}_S\cdot\mathbf{F}^s_i,
\end{split}
\end{equation} 
where the antisymmetric tensor $\mathbf{L}_A$ and the symmetric traceless tensor $\mathbf{L}_S$ are given by

\begin{equation}
\begin{split}
&L_{A,\alpha\beta}=\epsilon_{\alpha\gamma\beta}d_{\gamma},\\
&L_{S,\alpha\mu\nu}=\overbracket{d_{\mu}\delta_{\alpha\nu}} {}^{\substack{(\mu\nu) \ {} \\ {}}}.
\end{split}
\label{TensorL}
\end{equation}
%
Analogously, we can express the velocity multipoles of the rod in terms of the single particle velocity multipoles. Then, comparing with Eq. (\ref{Rod_friction}), we obtain the following approximate formulae for the friction tensors:

\begin{equation}
\begin{split}
&\mathbf{A}=\sum_{ij}\boldsymbol{\zeta}_{ij}\\
&\mathbf{C}=D^2\sum_{ij}ij\mathbf{L}_A\cdot\boldsymbol{\zeta}_{ij}\cdot
\mathbf{L}^T_A\\
&\mathbf{H}=D^2\sum_{ij}ij\mathbf{L}_S\cdot\boldsymbol{\zeta}_{ij}\cdot
\mathbf{L}^T_A\\
&\mathbf{M}=D^2\sum_{ij}ij\mathbf{L}_S\cdot\boldsymbol{\zeta}_{ij}\cdot
\mathbf{L}^T_S.
\end{split}
\label{Tensors2}
\end{equation}
Inserting Eqs. (\ref{TensorL}) and (\ref{Zeta_ij}) into Eq. (\ref{Tensors2}) and comparing with Eq. (\ref{Tensors}), we obtain

\begin{align}
&X^A=\zeta\sum_{ij}\phi_{ij},\\
&Y^A=\zeta\sum_{ij}\psi_{ij},\\
&X^M=\zeta D^2\sum_{ij}ij\phi_{ij},\\
\label{XMsum}
&Y^M=\frac{1}{2}Y^H=\frac{1}{2}Y^C=\frac{1}{2}\zeta D^2\sum_{ij}ij\psi_{ij},\\
&Z^M=0.
\end{align}
The coefficient $Z^M$ in the Oseen approximation equals zero, because it corresponds to a velocity field which vanishes along the rod. Inserting the above expressions into Eq. (\ref{Widehat_M}), the tensor $\widehat{\mathbf{M}}$ reduces to  

\begin{equation}
\widehat{\mathbf{M}}=\zeta D^2\sum_{ij}ij\phi_{ij}\mathbf{d}^{(0)}.
\label{Widehat_M_as}
\end{equation}
We note that the absence of the terms proportional to $\mathbf{d}^{(1)}$ and $\mathbf{d}^{(2)}$ for very long rods has a clear physical interpretation. It can be shown\cite{Kim_Karilla} that any linear, symmetric and traceless tensor $\mathbf{E}$ can be written as a sum $\sum_{i=0,1,2}\mathbf{d}^{(i)}:\mathbf{E}$. However, only the rate of strain of the form $\mathbf{d}^{(0)}:\mathbf{E}$ stretches the rod along its axis not producing a vorticity. Hence, the absence of terms proportional to $\mathbf{d}^{(1)}$ and $\mathbf{d}^{(2)}$ means that effectively, a very long, freely moving rod immersed in an arbitrary linear field disturbs the fluid flow such, as if it was immersed just in a field of a rate of strain $\mathbf{d}^{(0)}:\mathbf{E}$.

Here, we indicate a correspondence between the above expression and the general formula (\ref{Mu_general}). Namely, the Oseen tensors $\mathbf{T}_{0i}$ in Eq. (\ref{Mu_general}), from the summation rules (\ref{Summation}), can be replaced by $\mathbf{T}_{0i}-\mathbf{T}(\mathbf{R})$, which for large distances becomes $iD\mathbf{d}\cdot\nabla\mathbf{T}$. Then, again according to Eqs. (\ref{Summation}), we are left with $\boldsymbol{\mu}_{00}^{(2)}$ proportional to $\sum_{ij}ij\phi_{ij}$, in agreement with Eq. (\ref{Widehat_M_as}).

The two--particle mobility trace $\text{Tr}[\boldsymbol{\mu}_{00}^{(2)}]$ for large distances $R$ can be obtained from Eq. (\ref{Mobility_asymptotic}). After performing the trace and calculating the double sums in Eq. (\ref{Widehat_M_as}) using Eqs. (\ref{f_phi}) and (\ref{Fxy}), we arrive at

\begin{equation}
\text{Tr}[\boldsymbol{\mu}_{00}^{(2)}]=-\dfrac{\mu_0p^3}{128\mathcal{R}^4\log p} [1-6\cos^2\theta+9\cos^4\theta]\cdot[1+O((\log p)^{-1})],
\label{Trace_asymptotic}
\end{equation}
where $\mathcal{R}=R/D$. Comparing with (\ref{Tr_fit}) we find

\begin{equation}
		\begin{split}
		A  &\xrightarrow[p\rightarrow\infty]{}\frac{p^3}{128 \log p}\\
		C_1&\xrightarrow[p\rightarrow\infty]{}-6\\
		C_2&\xrightarrow[p\rightarrow\infty]{}9.
	\end{split}
\end{equation}

\bibliographystyle{phaip}

\begin{table}
\caption{
The first virial coefficient $\alpha$ as a function of rod aspect ratio $p$ for different hydrodynamic rod models ($A$,$B$) and truncation orders $L_{max}$.}
	\begin{tabular}{@{}c@{}c@{}c@{}c@{}c@{}}
	\hline\hline
	\makebox[0.2\textwidth][c]{$p$} & \multicolumn{4}{@{}r@{}}{
	\begin{tabular}{c}
	\makebox[0.8\textwidth][c]{$\alpha$}\\
	\hline
	\end{tabular}
	}\\
	    & \makebox[0.2\textwidth][c]{$A$}  & \makebox[0.2\textwidth][c]{$A$}  & \makebox[0.2\textwidth][c]{$B$}       & \makebox[0.2\textwidth][c]{$B$}\\
	    & $L_{max}=1$ & $L_{max}=3$ &  $L_{max}=1$ &  $L_{max}=3$\\
	\hline
1 & 1.07 & 1.83 & 1.07 & 1.83\\
2 & 1.33 & 1.60 & 1.64 & 1.68\\
3 & 1.46 & 1.65 & \ldots   & \ldots  \\
4 & 1.63 & 1.79 & 2.01 & 1.88\\
6 & 2.03 & 2.14 & 2.44 & 2.26\\
8 & 2.44 & 2.54 & 2.87 & 2.68\\
10 & 2.84 & 2.93 & 3.30 & 3.09\\
12 & 3.24 & 3.43 & 3.71 & 3.50\\
14 & 3.63 & 3.72 & 4.11 & 3.90\\
16 & 4.01 & 4.11 & 4.50 & 4.29\\
20 & 4.75 & 4.85 & 5.26 & 5.06\\
25 & 5.64 & 5.75 & 6.18 & 5.98\\
30 & 6.48 & 6.61 & \ldots   & \ldots  \\
40 & 8.10 & 8.25 & 8.70 & \ldots  \\ 
60 & 11.1 & 11.3 & 11.8 & \ldots \\ 
80 & 13.9 & 14.1 & 14.6 & \ldots \\
100 & 16.5 & 16.8 & \ldots & \ldots\\
120 & 19.1 & 19.4 & \ldots & \ldots\\
150 & 22.7 & 23.1 & \ldots & \ldots\\
200 & 28.5 & \ldots & \ldots & \ldots\\
300 & 39.3 & \ldots & \ldots & \ldots\\
500 & 59.3 & \ldots & \ldots & \ldots\\
1000 & 104.3 & \ldots & \ldots & \ldots\\
\hline\hline
\end{tabular}
\label{Alpha_table}
\end{table}



\end{document}